\documentclass[12pt,bibliography=totoc,DIV=17]{scrartcl}
\pdfoutput=1
\usepackage[utf8]{inputenc}

\usepackage{amsmath,amsfonts,amssymb,dsfont,mathtools}
\usepackage{array}
\usepackage[symbol]{footmisc}
\usepackage{stackrel}
\let\refeq\undefined
\usepackage[mathscr]{euscript}
\usepackage{enumitem}
\usepackage{longtable}
\usepackage{booktabs}
\usepackage{multirow}
\usepackage{cite}
\usepackage{graphicx, ulem, hhline}
\usepackage[table,dvipsnames]{xcolor}
\usepackage{xspace}
\usepackage{needspace}
\usepackage{siunitx}
\usepackage{soul}
\usepackage{wasysym}              
\usepackage{hyphenat}             

\usepackage{afterpage}
\usepackage{etoolbox,xstring,xspace,calc,xifthen}
\usepackage{centernot}

\usepackage{tikz}
\usetikzlibrary{tikzmark, calc}

\tikzset{
    double color fill/.code 2 args={
        \pgfdeclareverticalshading[%
            tikz@axis@top,tikz@axis@middle,tikz@axis@bottom%
        ]{diagonalfill}{100bp}{%
            color(0bp)=(tikz@axis@bottom);
            color(50bp)=(tikz@axis@bottom);
            color(50bp)=(tikz@axis@middle);
            color(50bp)=(tikz@axis@top);
            color(100bp)=(tikz@axis@top)
        }
        \tikzset{shade, left color=#1, right color=#2, shading=diagonalfill}
    }
}
\tikzset{
    stripes fill/.code 2 args={
        \pgfdeclareverticalshading[%
            tikz@axis@top,tikz@axis@middle,tikz@axis@bottom%
        ]{diagonalfill}{100bp}{%
            color(0bp)=(tikz@axis@bottom);
            color(12.5bp)=(tikz@axis@bottom);
            color(25bp)=(tikz@axis@bottom);
            color(25bp)=(tikz@axis@top);
            color(37.5bp)=(tikz@axis@top);
            color(37.5bp)=(tikz@axis@bottom!99);
            color(50bp)=(tikz@axis@bottom!99);
            color(50bp)=(tikz@axis@top!99);
            color(62.5bp)=(tikz@axis@top!99);
            color(62.5bp)=(tikz@axis@bottom);
            color(75bp)=(tikz@axis@bottom);
            color(75bp)=(tikz@axis@top);
            color(100bp)=(tikz@axis@top)
        }
        \tikzset{shade, left color=#1, right color=#2, shading=diagonalfill}
    }
}
\newcommand\BGcell[4][0pt]{%
  \begin{tikzpicture}[overlay,remember picture]%
    \path[#4] ( $ (pic cs:#2) + (-.05\tabcolsep,2.95ex) $ ) rectangle ( $ (pic cs:#3) + (1.65\tabcolsep,-#1*\baselineskip-1.3ex) $ );%
  \end{tikzpicture}%
}%
\newcounter{BGnum}
\newcommand\cellBG[6]{%
    \multicolumn{#2}{#3%
        !{\BGcell{startBG\arabic{BGnum}}{endBG\arabic{BGnum}}{#1}%
            \tikzmark{startBG\arabic{BGnum}}%
         }%
            #4@{\!\!}%
        !{%
          \tikzmark{endBG\arabic{BGnum}}%
        }%
        !{\hspace{1.15ex}}#5!{\hspace{-1.15ex}}}%
        {#6}%
      \addtocounter{BGnum}{1}%
}
\newcommand\cellBGend[6]{%
    \multicolumn{#2}{#3%
        !{\BGcellend{startBG\arabic{BGnum}}{endBG\arabic{BGnum}}{#1}%
            \tikzmark{startBG\arabic{BGnum}}%
         }%
            #4@{\!\!}%
        !{%
          \tikzmark{endBG\arabic{BGnum}}%
        }%
        !{\hspace{.4ex}}#5!{\hspace{-.4ex}}}%
        {#6\hphantom{ }}%
      \addtocounter{BGnum}{1}%
}
\newcommand\BGcellend[4][0pt]{%
  \begin{tikzpicture}[overlay,remember picture]%
    \path[#4] ( $ (pic cs:#2) + (-.05\tabcolsep,2.95ex) $ ) rectangle ( $ (pic cs:#3) + (1.1\tabcolsep,-#1*\baselineskip-1.3ex) $ );%
  \end{tikzpicture}%
}%

\makeatletter
\newif\iftag@here
\newcommand*{\taghere}[1][0pt]
{\ifmeasuring@\else
  \global\tag@heretrue
  \tikz[remember picture,overlay]{\coordinate (taghere) at (0pt,#1);}%
\fi}

\def\place@tag{%
    \iftagsleft@
      \kern-\tagshift@
      \iftag@here
        \global\tag@herefalse
        \tikz[remember picture,overlay]%
          {\path (taghere) -| node[anchor=base]{\rlap{\boxz@}} (0pt,0pt);}%
      \else
        \if1\shift@tag\row@\relax
            \rlap{\vbox{%
                \normalbaselines
                \boxz@
                \vbox to\lineht@{}%
                \raise@tag
            }}%
        \else
            \rlap{\boxz@}%
        \fi
        \kern\displaywidth@
      \fi
    \else
      \kern-\tagshift@
      \iftag@here
        \global\tag@herefalse
        \tikz[remember picture,overlay]%
          {\path  (taghere) -|  node[anchor=base]{\llap{\boxz@}} (0pt,0pt);}%
      \else
        \if1\shift@tag\row@\relax
            \llap{\vtop{%
                \raise@tag
                \normalbaselines
                \setbox\@ne\null
                \dp\@ne\lineht@
                \box\@ne
                \boxz@
            }}%
        \else \llap{\boxz@}%
        \fi
      \fi
    \fi
}
\makeatother

\def\thefootnote{\arabic{footnote}}

\makeatletter
\newlength{\fnhskip}
\renewcommand\@makefntext[1]{
  \settowidth{\fnhskip}{\@makefnmark}
  \leftskip=\fnhskip
  \hskip-\fnhskip
  \@makefnmark#1
}
\makeatother

\renewenvironment{subequations}[1][]{
  \refstepcounter{equation}%
  \setcounter{parentequation}{\value{equation}}
  \setcounter{equation}{0}
  \def\theequation{\theparentequation\alph{equation}}%
  \let\parentlabel\label
  \ifx\\#1\\\relax\else\label{#1}\fi
  \ignorespaces
}{%
  \setcounter{equation}{\value{parentequation}}
  \ignorespacesafterend
}

\newcommand*{\nextParentEquation}[1][]{
  \refstepcounter{parentequation}
  \setcounter{equation}{0}
  \ifx\\#1\\\relax\else\parentlabel{#1}\fi
}

\usepackage{setspace}

\usepackage[numbers,sort&compress]{natbib}
\makeatletter
\def\NAT@spacechar{\,}
\makeatother
\usepackage[
colorlinks=true
,urlcolor=blue
,anchorcolor=blue
,citecolor=blue
,filecolor=blue
,linkcolor=blue
,menucolor=blue
,linktoc=all
,unicode
,backref=page
]{hyperref}
\usepackage{hypernat}

\newrobustcmd*{\tocref}[1]{\hyperref[TOC]{\color{black}{#1}}}
\usepackage[all]{hypcap}
\usepackage{footnotebackref}

\usepackage[format=plain, margin=0.5cm, font=footnotesize, labelfont=bf, list=off]{caption}

\renewcommand*{\backref}[1]{}
\renewcommand*{\backrefalt}[4]{%
  \ifcase #1%
  \or [p\,#2]%
  \else [pp\,#2]%
  \fi%
}

\newif\ifbackrefshowonlyfirst
\backrefshowonlyfirsttrue
\makeatletter
\let\BR@direct@old@hyper@natlinkstart\hyper@natlinkstart
\renewcommand*{\hyper@natlinkstart}{\phantomsection\BR@direct@old@hyper@natlinkstart}
\let\BR@direct@oldBR@citex\BR@citex
\renewcommand*{\BR@citex}{\phantomsection\BR@direct@oldBR@citex}%

\long\def\hyper@page@BR@direct@ref#1#2#3{\hyperlink{#3}{#1}}

\ifx\backrefxxx\hyper@page@backref
    \let\backrefxxx\hyper@page@BR@direct@ref
    \ifbackrefshowonlyfirst
    \fi
\else
    \ifbackrefshowonlyfirst
    \fi
\fi

\patchcmd{\Hy@backout}{Doc-Start}{\@currentHref}{}{\errmessage{I can't seem to patch backref}}
\makeatother

\let\theparentequation\theequation
\patchcmd{\theparentequation}{equation}{parentequation}{}{}

\apptocmd{\thebibliography}{\scriptsize}{}{}
\let\OLDthebibliography\thebibliography
\renewcommand\thebibliography[1]{
  \OLDthebibliography{#1}
  \setlength{\parskip}{1pt}
  \setlength{\itemsep}{1pt plus 0.3ex}
}

\addtokomafont{disposition}{\rmfamily}
\BeforeTOCHead[toc]{{\pdfbookmark[1]{\contentsname}{toc}}}

\makeatletter
\patchcmd{\upbracefill}{\m@th}{\scriptstyle\m@th}{}{}
\patchcmd{\upbracefill}{$\braceld$}{$\scriptstyle\braceld$}{}{}
\patchcmd{\upbracefill}{\bracelu}{\bracelu\mkern-1mu}{}{}
\patchcmd{\upbracefill}{\hfill\braceru}{\hfill\mkern-1mu\braceru}{}{}
\DeclareOldFontCommand{\rm}{\normalfont\rmfamily}{\mathrm}
\DeclareOldFontCommand{\sf}{\normalfont\sffamily}{\mathsf}
\DeclareOldFontCommand{\tt}{\normalfont\ttfamily}{\mathtt}
\DeclareOldFontCommand{\bf}{\normalfont\bfseries}{\mathbf}
\DeclareOldFontCommand{\it}{\normalfont\itshape}{\mathit}
\makeatother

\newlength{\floatwidth}
\def\beq{\begin{equation}}
\def\eeq{\end{equation}}

\newcommand{\AtoB}[2]{\mbox{$#1\to #2$}}
\newcommand{\Amp}[4][\mathcal{A}]{#1^{\mbox{\tiny #2}}_{\mbox{\tiny #3}}\ifthenelse{\isempty{#4}}{}{{\left[#4\right]}}}

\def\twomat[#1,#2][#3,#4]{\left( \begin{array}{cc} #1 & #2 \\ #3 & #4 \end{array} \right)}
\def\threemat[#1,#2,#3][#4,#5,#6][#7,#8,#9]{\left( \begin{array}{ccc} #1 & #2 & #3\\ #4 & #5 & #6 \\ #7 & #8 & #9 \end{array} \right)}
\def\twovec[#1,#2]{\left( \begin{array}{c} #1  \\ #2 \end{array} \right)}
\def\thv[#1,#2,#3]{\left( \begin{array}{c} #1 \\ #2 \\ #3 \end{array} \right)}
\def\twv[#1,#2]{\left( \begin{array}{c} #1 \\ #2 \end{array} \right)}

\newcommand{\IE}{\textit{i.\,e.}\xspace}
\newcommand{\EG}{\textit{e.\,g.}\xspace}
\newcommand{\AP}{\mbox{\textit{a~priori}}\xspace}

\newcommand{\IL}{\textit{in~lieu}\xspace}
\newcommand{\refeq}[1]{Eq.\,\eqref{#1}}
\newcommand{\refeqs}[1]{Eqs.\,\eqref{#1}}
\newcommand{\sect}[1]{Sect.\,\ref{#1}}
\newcommand{\appx}[1]{Appx.\,\ref{#1}}
\newcommand{\fig}[1]{Fig.\,\ref{#1}}
\newcommand{\tab}[1]{Tab.\,\ref{#1}}
\newcommand{\citere}[1]{Ref.\,\cite{#1}}
\newcommand{\citeres}[1]{Refs.\,\cite{#1}}
\newcommand{\simord}{\mathord{\sim}\,}

\newcommand{\lsim}{\lesssim}
\newcommand{\gsim}{\gtrsim}
\newcommand{\lsimord}{\mathord{\lesssim}\,}
\newcommand{\gsimord}{\mathord{\gtrsim}\,}

\newcommand{\yint}[3]{Y_{#1}^{[#2]}\ifthenelse{\isempty{#3}}{}{{\left(#3\right)}}}
\newcommand{\ysum}[3]{Y_{#1}^{#2}\ifthenelse{\isempty{#3}}{}{{\left(#3\right)}}}
\newcommand{\yonesum}[3]{{^0}Y_{#1}^{#2}\ifthenelse{\isempty{#3}}{}{{\left(#3\right)}}}
\newcommand{\tint}[2]{\ifthenelse{\isempty{#2}}{T_{#1}}{T_{#1}^{\left|#2\right.}}}

\newcommand{\toneint}[2]{\ifthenelse{\isempty{#2}}{{^0}T_{#1}}{^0T^{\left|#2\right.}_{#1}}}

\newcommand{\yoneint}[3]{{^0}Y_{#1}^{[#2]}\ifthenelse{\isempty{#3}}{}{{\left(#3\right)}}}
\newcommand{\bint}[1]{B_{0}\ifthenelse{\isempty{#1}}{}{{\left(#1\right)}}}
\newcommand{\aint}[1]{A_{0}\ifthenelse{\isempty{#1}}{}{{\left(#1\right)}}}
\newcommand{\binteps}[2]{\ifthenelse{\isempty{#2}}{B_{0}^{\left|#1\right.}}{B_{0}^{\left|#1\right.}{\!\left(#2\right)}}}

\newcommand{\CP}{\ensuremath{\mathcal{CP}}\xspace}

\newcommand{\DR}{\ensuremath{\overline{\text{DR}}}\xspace}

\newcommand{\Real}[1]{\Re\hspace{-1pt}\mathfrak{e}{\left[#1\right]}}

\newcounter{notecount}
\hypersetup{ pdfauthor={Me, myself and I}
	,pdftitle={Curing tachyonic tree-level syndrome in NMSSM light singlet scenarios} ,pdfsubject={}
	,pdfkeywords={} }

\begin{document}
\newcommand*{\mytitle}[1]{%
  \parbox{\linewidth}{\setstretch{1.5}\centering\Large\textsc{\textbf{\boldmath #1}}}
}

\thispagestyle{empty}

\def\thefootnote{\fnsymbol{footnote}}

\begin{flushright}
  BONN-TH-2021-07\\
  TTK-21-33
\end{flushright}

\vfill

\begin{center}

\mytitle{Curing tachyonic tree-level syndrome\\
  in NMSSM light-singlet scenarios}

\vspace{1cm}

Florian Domingo$^{1}$\footnote{email: florian.domingo@csic.es}
and
Sebastian Pa{\ss}ehr$^{2}$\footnote{email: passehr@physik.rwth-aachen.de}

\vspace*{1cm}

\textsl{
$^1$Bethe Center for Theoretical Physics \&
Physikalisches Institut der Universit\"at Bonn,\\
Nu\ss allee 12, D--53115 Bonn, Germany
}

\medskip
$^2$\textsl{Institute for Theoretical Particle Physics and Cosmology,}\\
\textsl{RWTH Aachen University, Sommerfeldstra{\ss}e 16, 52074 Aachen, Germany.}

\end{center}

\vfill

\begin{abstract}{}
Models with an extended Higgs sector open up the phenomenological
possibility of additional scalars, beyond the SM-like boson observed
by the LHC, with mass at or below the electroweak scale. Such
scenarios are in particular viable in the presence of
electroweak-singlet spin-$0$ fields, as expected for instance in the
context of the~NMSSM. Given that the size of radiative corrections can
substantially affect the Higgs potential, a negative squared mass at
the tree level does not necessarily yield a tachyonic spectrum at the
physical level, but only indicates a failure of the tree-level
description for calculational purposes. We explain how to tackle this
technical issue in the example of the~NMSSM, in scenarios with light
\CP-odd or \CP-even singlet-dominated states and show how loop
corrections to the Higgs masses and decay widths can be derived with
the regularized Lagrangian. We further explore how the same
flexibility in the definition of tree-level parameters can be
exploited to circumvent large deviations of the tree-level spectrum
from the kinematical setup in Higgs decays, or to estimate the
theoretical uncertainty associated with the discrepancy between
tree-level and physical Higgs spectra. The latter is of particular
relevance for the properties of the SM-like Higgs boson in
supersymmetry-inspired models.
\end{abstract}

\vfill
\def\thefootnote{\arabic{footnote}}
\setcounter{page}{0}
\setcounter{footnote}{0}
\newpage
\section{Introduction}

The investigation of the Higgs properties at the
LHC\,\cite{Khachatryan:2016vau,Sirunyan:2018koj,Aad:2019mbh} places
limits on the phenomenology of models with extended Higgs sectors,
which should accommodate a state with characteristics roughly
Standard-Model (SM)-like at
$125.25\pm0.17$\,GeV\,\cite{Aad:2015zhl,CMS:2020xrn,ATLAS:2018tdk,ParticleDataGroup:2020ssz}. While
such a condition (up to the exact mass value) is almost automatically
fulfilled in scenarios with decouplingly heavy new physics, it should
be examined with care in the presence of additional light
states. Direct searches of the latter, as well as observables in
electroweak (EW) or flavor physics, are of course relevant tests of
such configurations as well. In order to exploit the growing
collection of experimental limits and constrain the parameter space of
models beyond the SM (BSM) in a quantitative fashion, precision
calculations are needed on the theoretical side. As a result, the mass
of the SM-like state in supersymmetric (SUSY) extensions of the
SM\,\cite{Nilles:1983ge,Haber:1984rc} has been the object of
considerable study in the last few decades---we refer the reader to
the recent review\,\cite{Slavich:2020zjv} for an overview of the
corresponding literature. This one observable is however far from
sufficient for a survey of the properties of the Higgs sector at the
EW~scale in non-minimal models.

In this paper, we focus on the Next-to-Minimal SUSY SM
(NMSSM)\,\cite{Ellwanger:2009dp,Maniatis:2009re}, and more
specifically on scenarios involving a light singlet-dominated state,
\IE~with mass below that of the observed SM-like Higgs boson, which is
one of the trademarks of Higgs physics in this particular
model\,\cite{Dobrescu:2000jt,Dobrescu:2000yn,Dermisek:2006wr,Morrissey:2008gm}. We
refer the curious reader to \EG~the comparatively recent works
concerning searches at colliders in
\citeres{Badziak:2016tzl,Das:2016eob,Muhlleitner:2017dkd,Guchait:2017ztk,Ellwanger:2017skc,Beskidt:2017dil,Basler:2018dac,Badziak:2018ijy,Beskidt:2019mos,Choi:2019yrv,Ma:2020mjz,Almarashi:2021blm}
as well as discussions in the context of singlino Dark Matter in
\citeres{Ellwanger:2018zxt,Domingo:2018ykx,Abdallah:2019znp,Wang:2020dtb,Barman:2020vzm,KumarBarman:2020ylm,Guchait:2020wqn,Abdallah:2020yag,Cao:2021tuh}. For
simplicity, we only consider the $R$-parity, $Z_3$- and \CP-conserving
version of the Lagrangian, but the methods and results that we develop
here can easily be extended to variants and even to distinct
models. We aim at extending our work on controlled predictions for the
Higgs masses and decays at the full one-loop~(1L) $\big/$ leading
two-loop~(2L) order in
the~NMSSM\,\cite{Domingo:2017rhb,Domingo:2018uim,Domingo:2019vit,Domingo:2020wiy,Domingo:2021kud}---similar
projects have also been presented in
\EG~\citeres{Graf:2012hh,Muhlleitner:2014vsa,Goodsell:2014pla,Goodsell:2015ira,Goodsell:2016udb,Goodsell:2019zfs,Dao:2021khm,Goodsell:2017pdq,Belanger:2017rgu,Baglio:2019nlc,Dao:2020dfb,Braathen:2021fyq}.

Our main interest in the present discussion consists in addressing a
technical difficulty that routinely arises in the presence of light
scalar states, namely the fact that the associated tree-level
squared-mass becomes negative while continuity suggests that the
physical squared mass (as estimated from the inclusion of radiative
corrections) would remain positive. As the evaluation of loop
functions with tachyonic spectrum is problematic and a standard
processing of the parameter points with negative mass-squared thus
fails, such benchmarks are often dismissed, while there is nothing
intrinsically wrong with them, only with the chosen description at
tree level. After a fashion, this issue is just an extension of the
one raised in the Coleman--Weinberg
mechanism\,\cite{Coleman:1973jx}. We address this `tachyonic
tree-level syndrome' in the context of the NMSSM in scenarios
involving light singlet-dominated \CP-even or \CP-odd states. We
further investigate the light-Higgs decays and the properties of the
SM-like state for these scenarios and tackle analogous problems caused
by the discrepancy between the tree-level and the physical spectrum at
the level of the three-point functions. We also insist on the accuracy
that can be achieved in such calculations at four-loop~(4L)~QCD
and~1L~EW/SUSY order: the latter can considerably degrade in the
presence of light singlet-dominated \CP-even states, limiting the
testability of corresponding scenarios at colliders.

In \sect{sec:mass} we describe our strategy to circumvent the
immediate difficulty of an artificial tachyonic spectrum at tree level
and illustrate the workings of this method over a few scenarios
involving light \CP-odd or \CP-even singlet-dominated states. The
decays of the light-Higgs state are investigated in
\sect{sec:singletdec}: in particular, we insist on the difficulty to
accurately assess the singlet--doublet mixing in 1L~calculations in
scenarios with a light \CP-even singlet. The phenomenology of the
SM-like state is discussed in \sect{sec:SMHiggs}, where we highlight
the impact of various sources of theoretical uncertainty in BSM
theories. Both SM-like and NMSSM-specific decay channels are
considered at this level. The achievements of the paper are
\mbox{summarized in the conclusions of \sect{sec:concl}}.

\section{The `tachyonic tree-level syndrome' and its cure \label{sec:mass}}

In models with extended Higgs sectors, in particular the NMSSM,
several Higgs states may take a mass at the EW scale or below without
necessarily endangering the limits from collider searches. However, as
the hierarchies in the spectrum generate large logarithms in loop
effects, radiative corrections to the masses may be comparable to the
tree-level values (or larger), raising the question of how to
interpret a tachyonic tree-level mass: does it signal an instability
of the selected minimum of the scalar potential, incompatible with the
phenomenology? Or is it just an artifact of the tree-level description
that radiative corrections can mend? We investigate this issue below
over several examples borrowed to the NMSSM phenomenology.

\subsection{Scenarios with a light \CP-odd Higgs\label{sec:CPoddmass}}

To state the issue in a clear fashion, we first turn to NMSSM
configurations where a light \CP-odd Higgs emerges. We remind the form
of the tree-level mass matrix for the \CP-odd Higgs sector (after
rotating out the Goldstone boson):\footnote{For our notations and
  conventions, we refer the reader to
  \citeres{Ellwanger:2009dp,Domingo:2017rhb}. The
    sets of \CP-even fields~$\{h_1, h_2, h_3\}$ and \CP-odd
    fields~$\{h_4, h_5\}$ are separately sorted by increasing mass. In
    addition, we define the symbols~$h^{\text{SM}}$, $h^{S}$
    and~$a^{S}$ for the SM-like, \CP-even singlet-dominated and
    \CP-odd singlet-dominated states, respectively.}
\begin{subequations}\label{eq:CPoddmat}
\begin{align}
{\cal M}_{\text{odd},11}^{2} &= M^2_A \equiv M_{H^{\pm}}^2-M_W^2+\lambda^2\,v^2\,,\\
{\cal M}_{\text{odd},22}^{2} &=
  M_A^2\left(\frac{\lambda\, v}{2\,\mu_{\text{eff}}}\,s_{2\beta}\right)^2
  - 3\,\frac{\kappa}{\lambda}\,\mu_{\text{eff}}\,A_{\kappa}
  + \frac{3}{2}\,\kappa\,\lambda\,v^2\,s_{2\beta}\,,\label{eq:odd22}\\
{\cal M}_{\text{odd},12}^{2} &= M_A^2\,\frac{\lambda\, v}{2\,\mu_{\text{eff}}}\,
  s_{2\beta} - 3\,\kappa\,v\,\mu_{\text{eff}}\,.
\end{align}
\end{subequations}
While the doublet component (${\cal M}_{\text{odd},11}^{2}$) is
typically heavy in realistic scenarios, due to constraints on the
\CP-even and charged sectors, see
\EG~\citeres{
  CMS:2018rmh,CMS:2018amk,ATLAS:2018gfm,CMS:2019bfg,ATLAS:2020zms,CMS:2020osd},
the singlet entry (${\cal M}_{\text{odd},22}^{2}$) can take values at
the EW~scale and below---even close to zero---in a phenomenologically
more realistic way because of the elusive nature of singlet-dominated
matter at colliders. Technically, a low-mass value easily emerges
when~$A_{\kappa}\!\to0$, with two more motivated limits where the
light pseudo-scalar can be viewed as the pseudo-Goldstone boson of an
approximate global $U(1)$~symmetry.

We first consider an NMSSM scenario corresponding to the
Peccei--Quinn~(PQ) limit of the model,
with~$\lvert\kappa/\lambda\rvert\ll1$: $\kappa=-0.02$, $\lambda=0.4$,
$M_{H^{\pm}}\!=2.5$\,TeV, $t_{\beta}=5$,
$\mu_{\text{eff}}\approx-480$\,GeV and we
vary~$A_{\kappa}\in[-60,60]$\,GeV. Then, terms violating the
$U(1)_{\text{PQ}}$~symmetry---with charge
assignment~$Q^{\text{PQ}}_S+Q^{\text{PQ}}_{H_u}+Q^{\text{PQ}}_{H_d}\stackrel[]{!}{=}0$
for the singlet~$S$ and doublet-Higgs fields~$H_{u,d}$---are
suppressed in the Higgs potential and a comparatively light
singlet-dominated pseudo-scalar Higgs state~\mbox{$a^S = h_4$} is thus
expected. With~$A_{\kappa}\!\gsim-6$\,GeV, the
mass-squared~$m_{a^S}^2$ of this state becomes negative at the tree
level in the considered scenario.

Let us now turn to the radiative corrections. Our renormalization
scheme has been described in \citere{Domingo:2017rhb} and employs
vanishing tadpoles, on-shell~(OS) conditions for SM-fermion and
EW-gauge-boson masses, as well as~$M_{H^{\pm}}^2$, while remaining
parameters are renormalized in the $\overline{\text{DR}}$~scheme, with
renormalization scale~$\mu_{\text{ren}}\!=m_t$. In the pseudo-scalar
Higgs sector, we restrict ourselves to corrections of 1L~order (since
those are currently the only fully exploitable radiative corrections
for a singlet-dominated \CP-odd state---see however
\citere{Goodsell:2014pla} as well as the recent \citere{Dao:2021khm}),
and evaluate them with the expansion-and-truncation method at
fixed-order described in \citeres{Domingo:2020wiy,Domingo:2021kud},
thus avoiding dependence of observable quantities on gauge-fixing
parameters and on the chosen field~renormalization. The Feynman
diagrams are computed with the help of
\texttt{FeynArts}\,\cite{Kublbeck:1990xc} and
\texttt{FormCalc}\,\cite{Hahn:2000kx,Hahn:1998yk}. Loop functions are
evaluated with \texttt{LoopTools}\,\cite{Hahn:1998yk}. The
doublet-dominated pseudo-scalar is evidently very far in mass and the
mixing of the singlet-dominated state with the EW~neutral~current
proves numerically negligible, justifying the use of a non-degenerate
framework for the derivation of mass corrections. As the SUSY~spectrum
becomes relevant at the radiative level, we specify that the squarks
of third generation take mass at~$\simord1.5$\,TeV, while~$M_3=2$\,TeV
and~$A_t\sim-2.5$\,TeV---for convenience, the input of all the
scenarios that we investigate is collected in \appx{app:in}. However,
for the singlet-dominated states, 1L~corrections primarily originate
in the Higgs and higgsino/singlino sectors. Then, the physical mass of
the light \CP-odd Higgs (as estimated at 1L~order)
for~$A_{\kappa}\lsim-6$\,GeV---\IE~in the
limit~$m^2_{a^S}\!\to0^+$---evaluates to~${\approx}\,70$\,GeV,
suggesting that the `tachyonic boundary' at~$A_{\kappa}\approx-6$\,GeV
is non-essential.

Parameter points with~$A_{\kappa}\gsim-6$\,GeV cannot be directly
processed in the formalism described above, because the derivation of
radiative corrections would require extending the definition of loop
functions for tachyonic masses (and external momenta). On the other
hand, such points cannot be dismissed on a robust phenomenological
basis and are just unexplorable in this setup. In view of the
technical issue at hand, one would be tempted to simply regularize the
tree-level spectrum by adding an ad-hoc contribution to the tree-level
mass-squared term, restoring its positivity, and then subtracting this
quantity at the radiative level: such a solution has been existing for
years within the public package
\texttt{NMSSMTools}\,\cite{Ellwanger:2004xm,Ellwanger:2005dv,Domingo:2015qaa},
which employs an effective-potential approach to the calculation of
the Higgs spectrum. Nevertheless, the explicit insertion of an
infrared~(IR) mass-cutoff in the lagrangian raises further concerns,
as such a term violates several symmetries at tree level, including
the EW~gauge~symmetry (since the light state contains a small doublet
component). In a `Feynman-diagrammatic' description taking into
account Higgs-to-Higgs mass corrections, it is therefore preferable to
address the question in a formally more robust fashion.

As we stated before, the issue actually originates in the tree-level
description of the parameter point, rather than in the scenario
itself. In other words, the tree-level parameters, \IE~the chosen
renormalization scheme, are just ill-suited for the description of
such scenarios, so that one simply needs to translate the problem to a
more convenient scheme. Numerous choices are \AP
possible. Nevertheless, in the $Z_3$-conserving NMSSM, a natural
handle on the masses of the singlet-dominated Higgs states is the
parameter~$A_{\kappa}$---see \refeq{eq:CPoddmat}. We thus shift its
tree-level value~$A_{\kappa}^{\text{orig}}\to
A_{\kappa}^{\text{mod}}\equiv
A_{\kappa}^{\text{orig}}+A_{\kappa}^{\text{shift}}$ and
correspondingly subtract the shift at the level of the
counterterm~$\delta A_{\kappa}^{\text{orig}}\to \delta
A_{\kappa}^{\text{mod}}\equiv\delta A_{\kappa}^{\text{orig}} -
A_{\kappa}^{\text{shift}}+\mathcal{O}(\text{2L})$, ensuring that we
are still describing the original point in parameter space (at the
perturbative order under control). The value
of~$A_{\kappa}^{\text{shift}}$ is chosen such that the tree-level
\CP-odd mass-squared in the modified scheme is
positive,~\mbox{$m^{2\,\text{mod}}_{a^S}>0$}. It is obvious from the
form of the mass matrix that (as long as~$\kappa\neq0$)\footnote{The
  case~$\kappa=0$ is phenomenologically problematic in the
  $Z_3$-conserving model, as the potential favors a very small vacuum
  expectation value (v.e.v.) for the singlet, leading to difficulties
  in view of the lower bound set by LEP~searches on chargino
  masses~$m_{\chi^\pm}\!\sim\mu_{\mathrm{eff}}=\lambda\,v_s$\,\cite{LEPSUSYWG}. In
  $Z_3$-violating versions, other parameters, \EG~the coefficients of
  tadpole or quadratic operators, lend themselves to comparable
  operations.}~$A_{\kappa}^{\text{shift}}$ can be chosen to
accommodate any value of~$m^{2\,\text{mod}}_{a^S}$. Nevertheless, we
also know from physical intuition that the farther this value stands
away from the physical mass, \IE~the pole of the propagator, the
slower the perturbative series converges, so that a wise choice
of~$m^{2\,\text{mod}}_{a^S}$ would select its value close to the
OS~scheme for the light singlet---for which, however, the conversion
factor from the original scheme cannot be determined without
regularizing the tree-level spectrum first. In
addition,~$A_{\kappa}^{\text{shift}}$ should remain of a magnitude
allowing its interpretation as a quantity of 1L~order---this can be
controlled with the requirement that large cancellations be absent
from the combination of tree-level and radiative contributions to the
mass of the light-singlet state.

\begin{figure}[t!]
  \centering\captionsetup{singlelinecheck=off}
  \includegraphics[width=\linewidth]{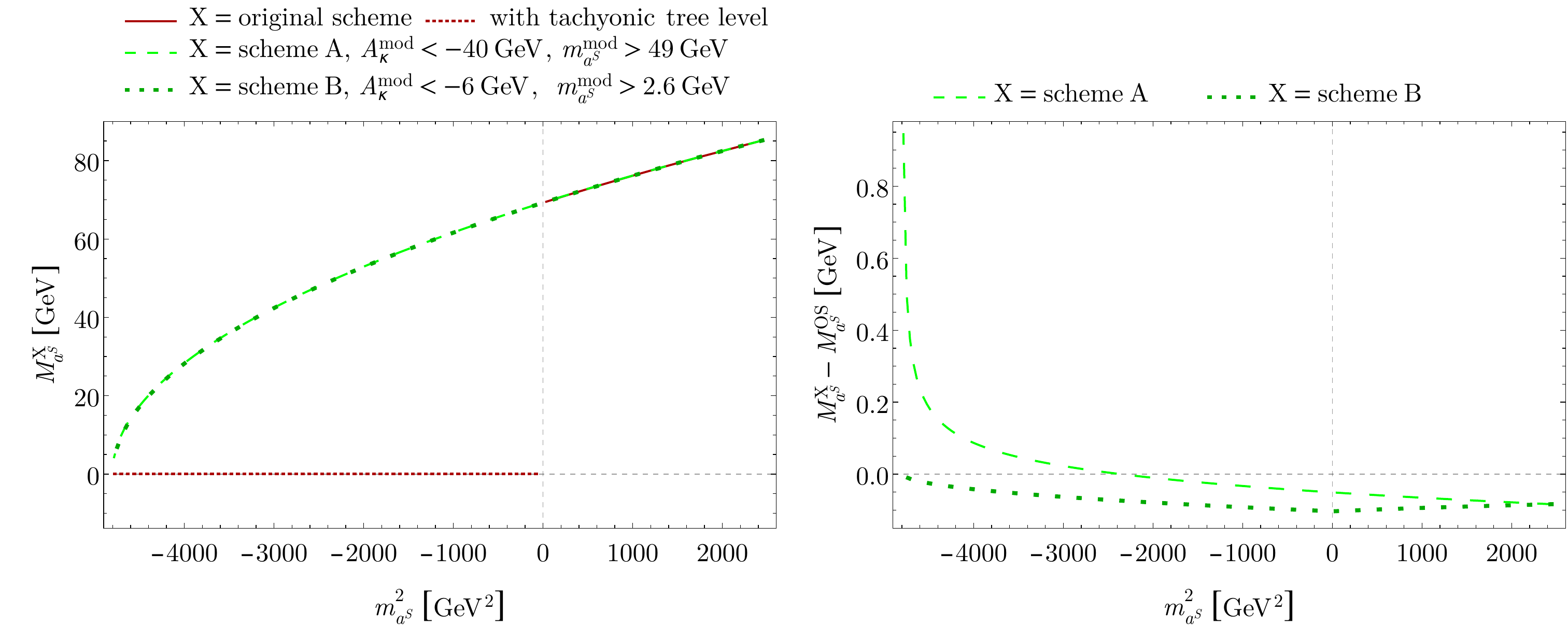}
  \caption{Mass predictions for the light \CP-odd Higgs~$a^S$ in the
    scenario of the PQ~limit, plotted against the tree-level
    mass-squared of the original scheme.
    \begin{itemize}[leftmargin=\widthof{{\em Right}:}+\labelsep]
    \item[{\em Left}:] Mass predictions in the original scheme (red),
    and in two modified schemes with~$A_{\kappa}$ shifted so
    that~$m^2_{a^S}$ remains above~$49$\,GeV (scheme\,A, dashed) or
    above~$2.6$\,GeV (scheme\,B, dotted).
    \item[{\em Right}:] Differences in mass predictions between
    scheme\,A (dashed) or~B (dotted) with the OS~scheme. The
    predictions of scheme\,B and the original scheme essentially
    coincide for~$m^2_{a^S}>0$.
    \end{itemize}\label{fig:PQmass}}
\end{figure}

In \fig{fig:PQmass}, we show the predicted pseudo-scalar
mass~$M_{a^S}$ at~1L in the considered scenario as a function
of the tree-level mass-squared in the original renormalization
scheme,~$m^2_{a^S}$. The calculation in the original scheme (red
curve) is only defined for positive~$m^2_{a^S}$ (for illustrational
purpose, the loop-corrected mass is replaced by~$0$ for
negative~$m^2_{a^S}$). Two modified schemes~A and~B are considered in
green, which essentially implement an IR-cutoff on the tree-level
pseudo-scalar mass with~$m^{2\,\text{(A)}}_{a^S}>49$\,GeV (dashed
curve) and~$m^{2\,\text{(B)}}_{a^S}>2.6$\,GeV (dotted curve)---through
the manipulations on the renormalization of~$A_{\kappa}$ that are
described above. Both extend the validity of the mass prediction well
into the negative-$m^2_{a^S}$ region and largely coincide with one
another, despite the very different choice of IR-cutoff for the
pseudo-scalar mass.

In the plot on the right of \fig{fig:PQmass}, we show the difference
of the loop-corrected masses predicted in schemes~A and~B with respect
to the OS~scheme
(\IE~$A_\kappa^{\text{mod}}\stackrel[]{!}{=}A_{\kappa}^{\text{OS}}$
ensuring~$m_{a^S}^{\text{OS}}\stackrel[]{!}{=}M_{a^S}^{\text{OS}}$; in
practice, this condition is approximately fulfilled by means of an
iterative algorithm that converges to the exact value). This quantity
typically remains of order~$100$\,MeV in the considered scenario, and
exceeds this magnitude only for scheme\,A in the
limit~$M_{a^S}\!\to0$, where the corresponding tree-level
mass~$m_{a^S}^{\text{(A)}}$ is comparatively far from the physical
mass and the hierarchy as compared to the EW~scale becomes more
relevant. In the region with~$m^2_{a^S}\!>0$, scheme\,B (with a very
low IR-cutoff) hardly differs from the original scheme, while the
predictions of scheme\,A (with a higher cutoff) depart from those of
the original scheme by effects
of~$\mathcal{O}(10\text{\,MeV})$. Naively, we expect scheme\,A to
perform better in the regime with a heavier~$M_{a^S}$, while scheme\,B
should offer a better depiction of the physics situation
when~\mbox{$M_{a^S}\!=\mathcal{O}(1\text{\,GeV})$}. The OS~scheme
serves as our reference, but its definition requires some further
operations as compared to the other two. However, in this specific
scenario, all these schemes perform quite similarly. We stress that
perturbative corrections in all schemes are of limited usefulness in
the regime with tree-level mass~$m_{a^S}\lsim m_c$ (or more
essentially for~$M_{a^S}\!\lsim m_c$) where interactions with
strongly-interacting matter are not captured by the partonic
picture. In fact, the description at~\mbox{$q$--$\bar{q}$}~thresholds
already calls for a more detailed analysis---see
\EG~\citere{Domingo:2009tb} and references therein.

\begin{figure}[t!]
  \centering
  \includegraphics[width=\linewidth]{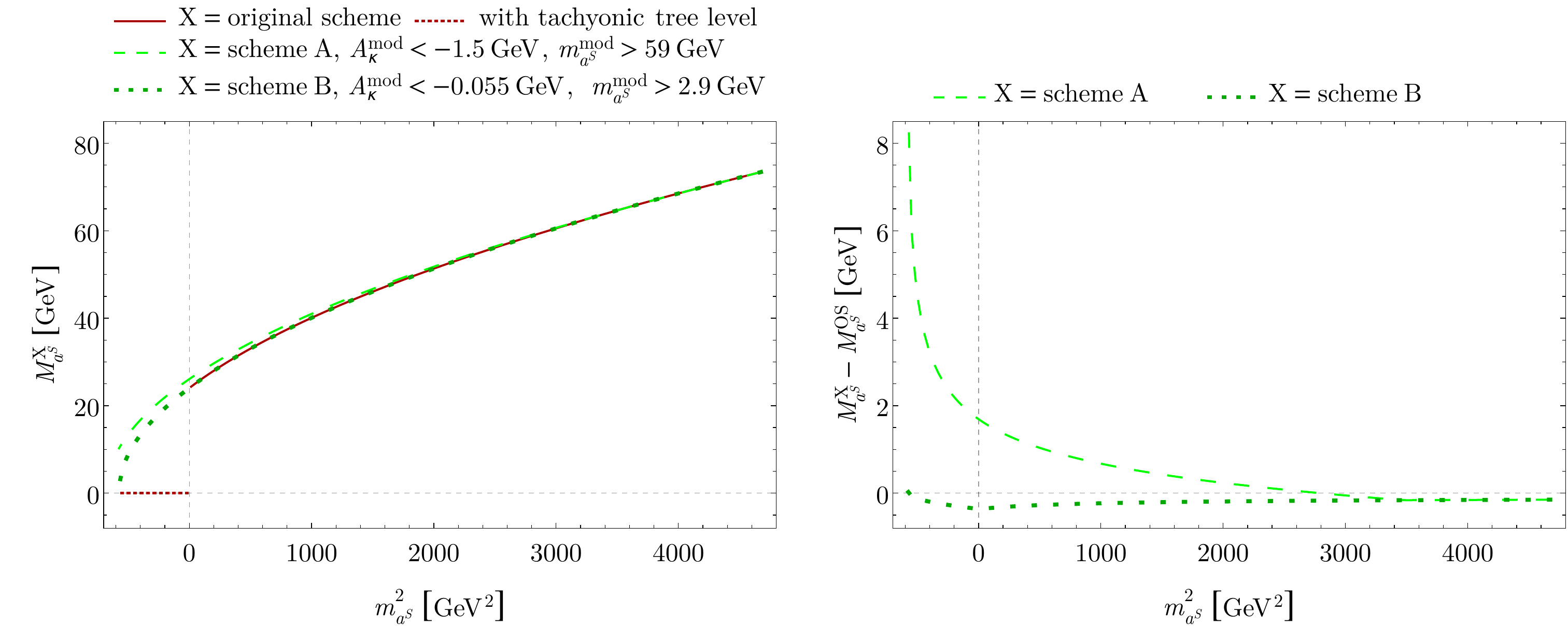}
  \caption{Mass predictions for the light \CP-odd Higgs~$a^S$ in the
    scenario of the $R$-symmetry limit, plotted against the tree-level
    mass-squared of the original scheme. The conventions are similar
    to those of \fig{fig:PQmass}, up to adjustments in the definition
    of the modified schemes. \label{fig:RSmass}}
\end{figure}

\needspace{5ex} We consider another example in \fig{fig:RSmass},
corresponding to the $R$-symmetry limit of the~NMSSM---\IE~with the
parametrization in \refeq{eq:CPoddmat}: $A_{\kappa}\to0$,
$M_A^2-2\,(\kappa/\lambda)\,\mu_{\text{eff}}^2\big/s^2_{2\beta}\to0$---where
again the light~$a^S = h_4$ can be viewed as the pseudo-Goldstone
boson of a global $U(1)_R$~symmetry with charge assignment
satisfying~$2\,Q^{R}_S-Q^{R}_{H_u}-Q^{R}_{H_d}\stackrel[]{!}{=}0$. Here,
$\kappa=0.4$, $\lambda=0.2$, $M_{H^{\pm}}\!=1.75$\,TeV,
$t_{\beta}=10$, $\mu_{\text{eff}}=400$\,GeV and we
vary~$A_{\kappa}\in[-2,0.5]$\,GeV. The tree-level mass of the lightest
\CP-odd Higgs becomes tachyonic for~$A_{\kappa}\gsim-0.05$\,GeV while
the 1L-corrected mass is still~${\approx}\,24$\,GeV. We again consider
two modified schemes~A and~B with IR-cutoffs at~$59$\,GeV
and~$2.9$\,GeV on the singlet-Higgs mass~$m_{a^S}$, respectively. In
this case, the mass predictions of scheme\,A (with a higher IR-cutoff)
depart from those of the OS~scheme (or those of scheme\,B) more
significantly than in the first example in the low-mass regime: the
difference reaches several~GeV; this is in part due to the choice of a
higher cutoff in scheme\,A, less suited to describe light states. In
addition, the non-negligible value of~$\kappa$ in the chosen benchmark
both generates a hierarchy in the singlet/singlino sector and makes
corresponding 1L~corrections to the pseudo-scalar mass sensitive to
this splitting. However, the qualitative agreement remains acceptable
and, in any case, the regularization always improves on the
unpredictive original scheme. In fact, these variations with the
choice of scheme also shed some light on the uncertainties at stake in
the mass predictions for the light Higgs states at the considered
order: it is indeed not surprising that, when the tree-level mass is
far from the physical one (in relative value), the convergence of the
perturbative series is slow, hence the theoretical uncertainty is
large at 1L~order.

Finally, a low pseudo-scalar Higgs mass can also be achieved
`accidentally', typically with a small value of~$A_{\kappa}$. The
Higgs/higgsino spectrum is then \AP~more free than in the~PQ- and
$R$-symmetry limit. We also stress that radiative corrections to the
squared mass of the light Higgs~$a^S$ are not systematically
positive, as in the examples above. Nevertheless, variations in the
scheme may still be of interest, \EG~in order to bring the tree-level
spectrum closer to the physical situation or to assess theoretical
uncertainties: the analysis of the Higgs decays will illustrate these
points in \sect{sec:singletdec} and \sect{sec:SMHiggs}.

\subsection{Scenarios with a light \CP-even Higgs\label{sec:CPevenmass}}

Another phenomenological possibility is that of a light \CP-even
state. We first remind the form of the \CP-even mass matrix in the
base of gauge eigenstates:
\begin{subequations}\label{eq:CPevenmat}\allowdisplaybreaks
\begin{align}
{\cal M}_{\text{even},11}^{2} &= M^2_A\,s_{\beta}^2+M_Z^2\,c_{\beta}^2\,,\\
{\cal M}_{\text{even},22}^{2} &= M^2_A\,c_{\beta}^2+M_Z^2\,s_{\beta}^2\,,\\
{\cal M}_{\text{even},33}^{2} &=
  M_A^2\left(\frac{\lambda\,v}{2\,\mu_{\text{eff}}}\,s_{2\beta}\right)^2
  + \frac{\kappa}{\lambda}\,\mu_{\text{eff}}
    \left(A_{\kappa}+4\,\frac{\kappa}{\lambda}\,\mu_{\text{eff}}\right)
  - \frac{1}{2}\,\kappa\,\lambda\, v^2\,s_{2\beta}\,,\label{eq:even33}\\
{\cal M}_{\text{even},12}^{2} &=
  -\frac{1}{2}\left(M_A^2+M_Z^2-2\,\lambda^2\,v^2\right) s_{2\beta}\,,\\
{\cal M}_{\text{even},13}^{2} &= \lambda\,v \left[2\,\mu_{\text{eff}}\,c_{\beta}
  - \left(\frac{M_A^2}{\mu_{\text{eff}}}\,s_{2\beta}
  + 2\,\frac{\kappa}{\lambda}\,\mu_{\text{eff}}\right) s_{\beta}\right],\\
{\cal M}_{\text{even},23}^{2} &= \lambda\,v \left[2\,\mu_{\text{eff}}\,s_{\beta}
  - \left(\frac{M_A^2}{\mu_{\text{eff}}}\,s_{2\beta}
  + 2\,\frac{\kappa}{\lambda}\,\mu_{\text{eff}}\right) c_{\beta}\right].
\end{align}
\end{subequations}
Once again, the singlet entry~${\cal M}_{\text{even},33}^{2}$ can be
small in the limit where the PQ-breaking scales~$A_{\kappa}$
and~$(\kappa/\lambda)\,\mu_{\text{eff}}$ are at the EW~scale or
below. In fact, in the PQ-inspired scenario of the previous section,
the spectrum contains a light \CP-even singlet-dominated state with
mass~$\simord50$\,GeV at tree level, and reaching~$90$--$100$\,GeV at
1L~order. In addition, an accidental cancellation between the terms
contributing to~${\cal M}_{\text{even},33}^{2}$ is possible, allowing
for a light \CP-even state without a light pseudo-scalar
simultaneously. In this case, as the low mass is not guaranteed by a
symmetry, radiative corrections are expected to have a sizable impact.

From the phenomenological perspective, a complication emerges in the
presence of a light \CP-even state, as its mixing with the SM-like
state should not excessively alter the properties of the latter: this
tends to imply either a very suppressed~$\lambda$ or a correlation
between~$\mu_{\text{eff}}$, $M_A$ and~$t_{\beta}$; the latter can be
deduced from the form of~${\cal M}_{\text{even},13}^{2}$ and~${\cal
  M}_{\text{even},23}^{2}$. As this relation is not guaranteed by a
symmetry, there is limited control over the singlet--doublet mixing at
the radiative level.

\begin{figure}[b!]
  \centering
  \includegraphics[width=\linewidth]{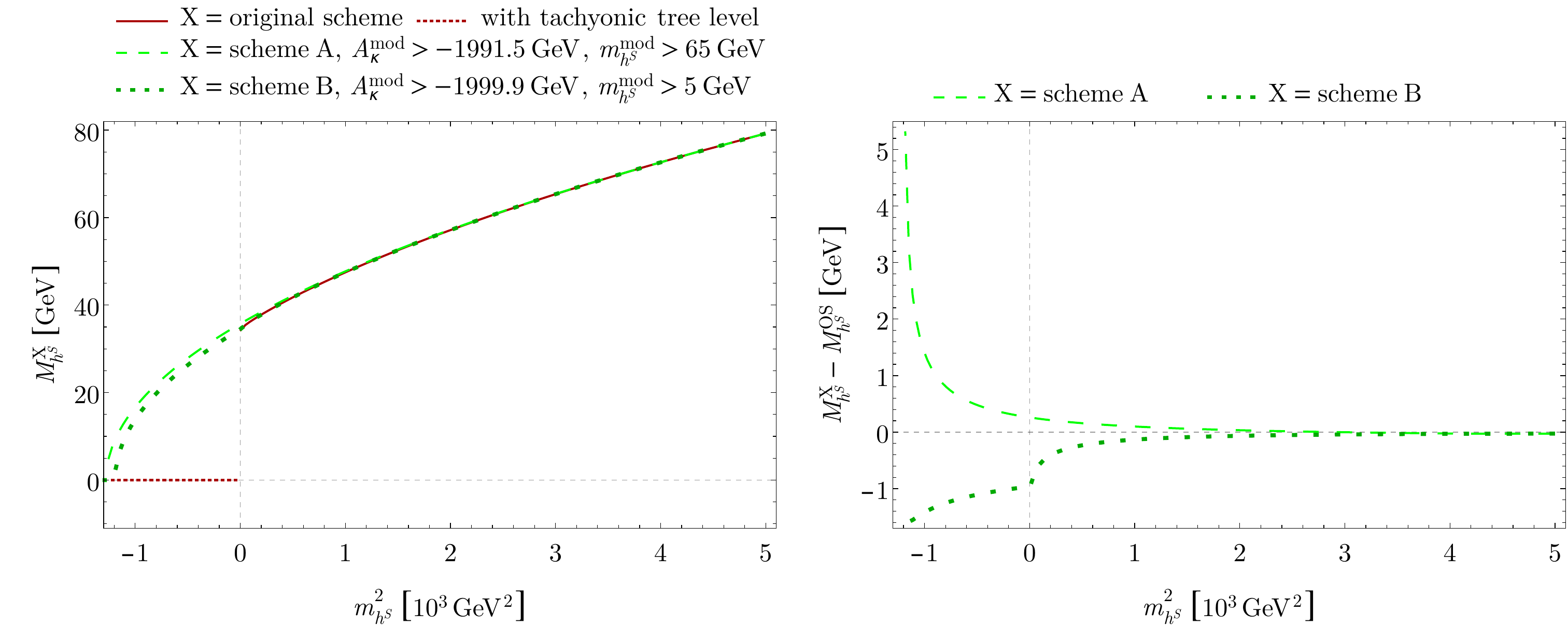}
  \caption{Mass predictions for the light \CP-even Higgs~$h^S$ in the
    scenario\,CPE1 with~$\lambda=\kappa=0.05$, plotted against the
    tree-level mass-squared of the original scheme. The conventions
    are similar to those of \fig{fig:PQmass}, up to adjustments in the
    definition of the modified schemes. \label{fig:CPEbismass}}
\end{figure}

We first consider a scenario with comparatively low values
of~$\lambda$ and~$\kappa$ (for commodity, we will refer to it as~CPE1
below): $\lambda=\kappa=0.05$, $M_{H^{\pm}}\!=4.8$\,TeV,
$t_{\beta}=10$, $\mu_{\text{eff}}=500$\,GeV and~$A_{\kappa}$ in the
range~$[-2003,-1990]$\,GeV; squark masses are at~$1.5$\,TeV and the
associated trilinear couplings at~$-2.4$\,TeV. The heavy-doublet and
\CP-odd singlet-dominated states then receive multi-TeV masses,
leaving two light \CP-even Higgs bosons: the SM-like~$h^{\text{SM}}=
h_2$ and the singlet-dominated~$h^S= h_1$. The latter two are largely
decoupled at the tree level, with mixing-squared at the level of
$10^{-4}$. For~$A_{\kappa}\lsim-2$\,TeV, the tree-level
mass-squared~$m_{h^S}^2$ of the light singlet becomes negative while
the predicted physical mass~$M_{h^S}$ is still
at~${\approx}\,34$\,GeV. Once again, the scheme of~$A_{\kappa}$ can be
adapted to perform the mass calculation in the pseudo-tachyonic
region, employing~$A_{\kappa}^{\text{shift}}$ to stabilize the
\CP-even singlet mass this time: this is shown in
\fig{fig:CPEbismass}. There, we again consider two schemes~A and~B
implementing an IR-cutoff on~$m_{h^S}$ at~$\simord65$\,GeV (scheme\,A)
or~$\simord5$\,GeV (scheme\,B), as well as the OS~scheme. All provide
qualitatively compatible results, although the low-mass regime shows
larger deviations.

The solution addressing the pseudo-tachyonic region is therefore
identical to the case of a light \CP-odd state, with the only
difference that the shift in~$A_{\kappa}$ is performed in the opposite
direction. In principle, a complication affecting this method is
possible in scenarios where both the \CP-odd and \CP-even
singlet-dominated states are very light (or tachyonic): then, the
shift of~$A_{\kappa}$ may not suffice to obtain positive tree-level
masses for both states simultaneously (the shift \AP~acts in opposite
directions for each, compare \refeqs{eq:odd22} and
\eqref{eq:even33}). Then, one may attempt to regularize the full
tree-level spectrum by also altering the renormalization scheme for
the other $Z_3$-conserving singlet
scale~$(\kappa/\lambda)\,\mu_{\mathrm{eff}}$. Alternatively, one may
simply add a quartic singlet coupling~$\hat{\kappa}^2\,\lvert
S\rvert^4$ to the Higgs potential as a handle on the \CP-even mass,
while~$A_{\kappa}$ is determined by the regularization of the \CP-odd
mass. We will not investigate such scenarios in detail below, although
we will demonstrate the use of the extended Higgs potential for the
PQ-inspired scenario of the previous subsection at the level of the Higgs
decays of the SM-like state.

\begin{figure}[b!]
  \centering
  \includegraphics[width=\linewidth]{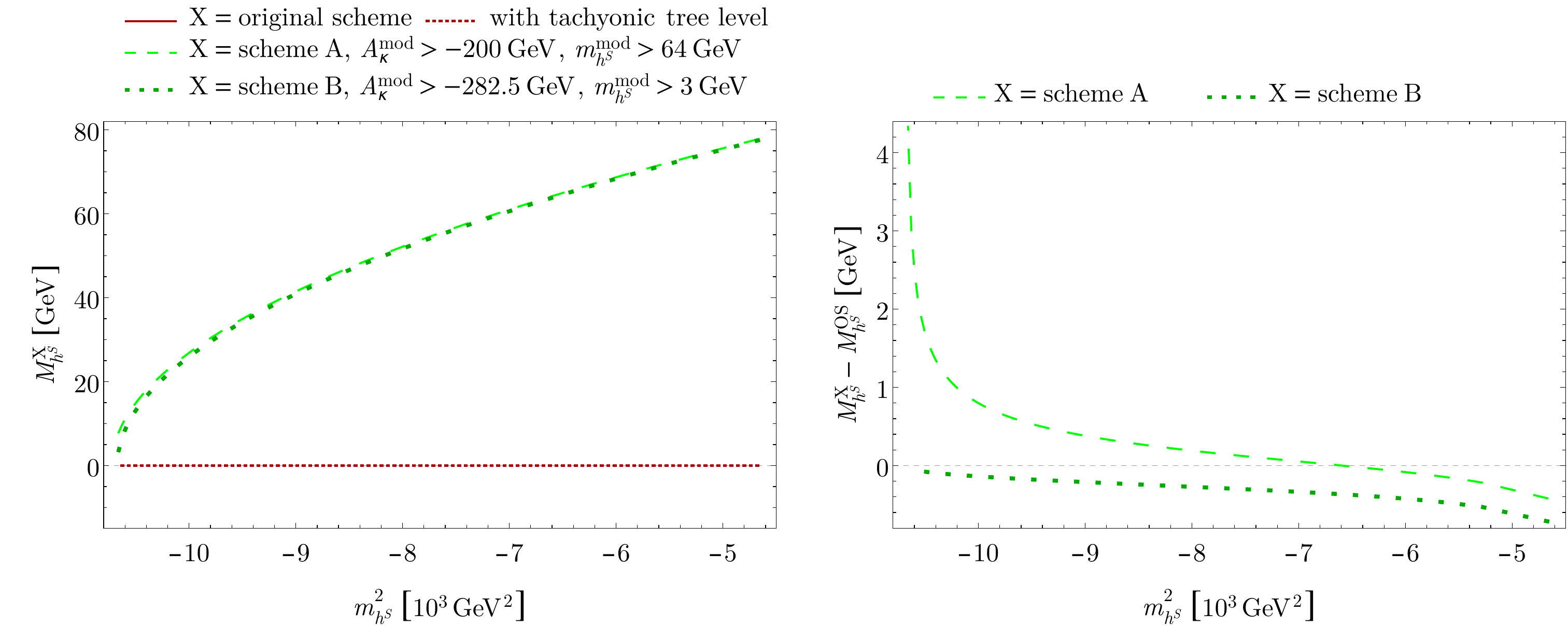}
  \caption{Mass predictions for the light \CP-even Higgs~$h^S$ in the
    scenario\,CPE2, plotted against the tree-level mass-squared of the
    original scheme. The conventions are similar to those of
    \fig{fig:PQmass}, up to adjustments in the definition of the
    modified schemes. \label{fig:CPEmass}}
\end{figure}

A second illustration is shown in \fig{fig:CPEmass} for a large value
of~$\lambda$ (the scenario is labelled~CPE2 below). In this case,
$\kappa=0.05$, $\lambda=0.5$, $M_{H^{\pm}}\!=1.225$\,TeV,
$t_{\beta}=2$, $\mu_{\text{eff}}=500$\,GeV and~$A_{\kappa}$ in the
range~$[-500,-380]$\,GeV. The soft-breaking sfermion mass parameters
are unchanged as compared to the earlier scenarios, but the trilinear
soft-breaking couplings are now set
to~$2.1$\,TeV. While~\mbox{$\kappa\ll\lambda$}, the scale of the
PQ-symmetry-breaking terms is comparatively high and the lightest
\CP-odd Higgs takes a mass above~$200$\,GeV. In the \CP-even sector,
the full range of~$A_{\kappa}$ leads to a tachyonic tree-level mass
for the singlet-dominated scalar~$h^S = h_1$: $m_{h^S}$~vanishes
at~$A_{\kappa}\approx-283$\,GeV, but the corresponding 1L-corrected
mass then reads~$M_{h^S}\approx103$\,GeV, demonstrating the large
magnitude of the radiative corrections and the artificial nature of
the tachyonic boundary at the tree level. In \fig{fig:CPEmass}, the
mass prediction at~1L is plotted in a scheme\,A with a high IR-cutoff
of~$65$\,GeV (dashed green) and a scheme\,B (dotted green) with a low
IR-cutoff of~$\simord3$\,GeV. The original scheme (red) does not
provide results in the considered range.

\section{Properties of the light Higgs at 1L order \label{sec:singletdec}}

The regularization of the tree-level spectrum through the redefinition
of the renormalization scheme further allows to compute the decay
properties of the light state at the radiative level. In fact, it
allows to tackle another issue appearing at the level of the
three-point functions: the tree-level Higgs mass may be smaller than
the sum of the masses of the final states (while the physical mass is
larger), thus causing problems with the evaluation of the loop
integrals with \texttt{LoopTools}.\footnote{\texttt{LoopTools} still
  provides a numerical result of dubious quality with copious
  warnings, eventually making the evaluation unstable.} This problem
is not fundamental in nature, but associated with the implemented
numerical method. It would thus be possible to repair the evaluation
of three-point functions to address all kinematical configurations. On
the other hand, we may worry about the accuracy of an evaluation that
is performed in a regime comparatively different from the actual one,
\IE~with tree-level Higgs masses significantly below the
  loop-corrected physical values, especially if it displaces the
external momenta with respect to poles of the internal propagators.
Employing a loop-corrected mass in the 1L~amplitudes is a
straightforward method for dealing with this issue without the need to
address the problematic regime, but it endangers the invariance of
observables with respect to gauge-fixing parameters and field
renormalization, as argued in \citere{Domingo:2020wiy}, hence
resulting in limited improvement in terms of accuracy. Therefore, it
is useful to bypass the problem in a more consistent way and compute
Higgs decays in a scheme that directly places the tree-level masses in
a kinematically allowed configuration: the OS~scheme appears as the
most convenient choice and also probably the most predictive, since it
brings the tree-level spectrum in accordance with the physical
configuration.

In view of the considered spectra and restricting ourselves to the
partonic picture, the accessible final states to the decays of the
light singlet-dominated state at~1L essentially consist of SM-fermion pairs at
  next-to-leading order~(NLO), as well as gluon and photon
pairs at leading order~(LO), including further radiation of
photons and gluons: we directly consider the inclusive fermionic decay
widths with respect to~QCD
and~QED\,\cite{Braaten:1980yq,Drees:1990dq}, including up to
$\mathcal{O}(\alpha_s^4)$
corrections\,\cite{Chetyrkin:1995pd,Baikov:2005rw}---see also the
short overview in \citere{Spira:2016ztx}---as well as QCD~corrections
to the diphoton~($\gamma\gamma$) and digluon~($gg$)
channels\,\cite{Spira:1995rr,Muhlleitner:2006wx}. Due to the hierarchy
of Yukawa couplings, the $b\bar{b}$~channel is expected to dominate if
kinematically accessible, with important decays into~$\tau^+\tau^-$,
$gg$ and~$c\bar{c}$ below the $b$--$\bar{b}$~threshold. Nevertheless,
such a partonic picture is over-simplistic at Higgs masses similar to
twice the bottom mass and below, and the interaction of the Higgs
state with hadronic matter should then be considered in a
non-perturbative fashion: we refer the reader to
\EG~\citeres{Drees:1989du,Domingo:2011rn} for a discussion of effects
at~$q$--$\bar{q}$~thresholds, as well as~\citere{Domingo:2016yih} for
the chiral limit. While we restrict ourselves to the partonic
description here, this should thus be seen only as the perturbative
step in a more involved path to the actual decay widths at low mass.

\subsection{Case of a light \CP-odd Higgs}

The decays of the light singlet-dominated
pseudo-scalar~$a^S = h_4$ into SM~matter proceed at tree
level through its subleading doublet component. The latter is
suppressed by the high mass of the doublet component,~$M_A\approx
M_{H^{\pm}}\!\gg M_Z$. The exact choice of~$M_A$ is not free in the
$R$-symmetry limit, as it is largely determined by the defining
condition of this limit,
\IE~\mbox{$M_A^2\approx2\,(\kappa/\lambda)\,\mu_{\text{eff}}^2\big/s^2_{2\beta}$}. A
less immediate condition also arises in the PQ-limit in order to avoid
large mixing in the \CP-even
sector:~$M_A^2\approx4\,\mu_{\text{eff}}^2\big/s^2_{2\beta}$. Correspondingly,
the size of the doublet component in~$a^S$ at tree level evaluates
to~${\approx}\,0.03$ and~${\approx}\,0.02$ in the two scenarios of
Sect.\,\ref{sec:CPoddmass}, leading to reduced decay widths (and a
suppressed direct production cross-section at colliders). Though
moderate in these examples, the value of~$t_{\beta}>1$ further favors
the decays into down-type quarks and leptons.

\begin{figure}[b!]
  \centering
  \includegraphics[width=\linewidth]{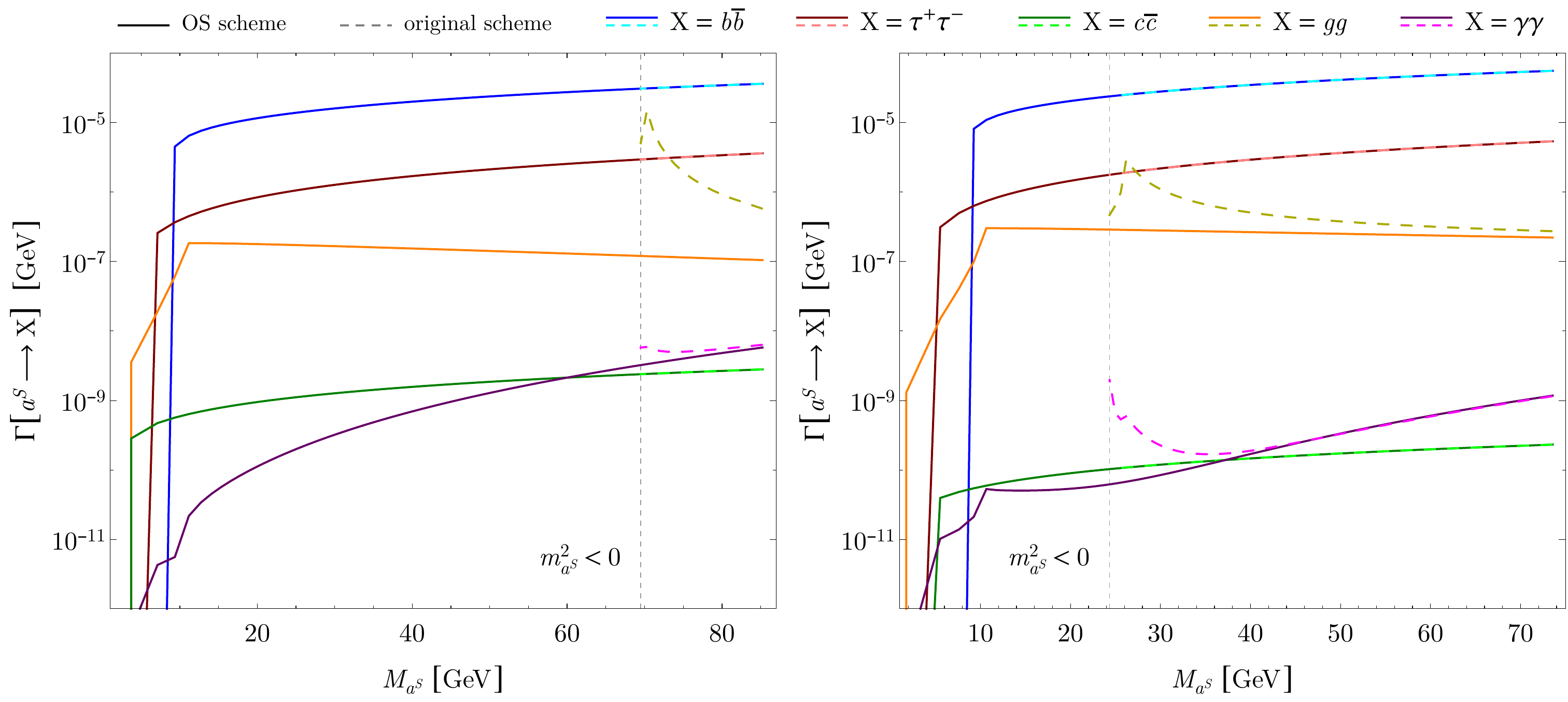}
  \caption{Partial decay widths of the light \CP-odd state~$a^S$ in
    the scenarios of Sect.\,\ref{sec:CPoddmass}: PQ- (left) and
    $R$-symmetry limit (right). The solid curves are obtained in the
    scheme with the renormalization of~$A_{\kappa}$ determined by the
    OS~condition on the mass of the light singlet~$a^S$. The decays in
    the original scheme are shown with dashed
    lines. \label{fig:A1dec}\vspace{-1.1ex}}
\end{figure}

The decay widths into~$b\bar{b}$, $\tau^+\tau^-$, $c\bar{c}$, $gg$
and~$\gamma\gamma$ are shown as a function of the physical Higgs mass
in \fig{fig:A1dec} for the two proposed benchmarks. Calculation in the
original scheme (dashed lines) not only fails when the tree-level mass
turns negative, but already when it falls under threshold (although
the physical mass is above), due to the aforementioned difficulty in
the evaluation of three-point loop functions with
\texttt{LoopTools}. Further spurious effects develop when the
tree-level mass is small (visible mostly in the digluon and diphoton
channels). Now allowing for a change of scheme via a shift
of~$A_{\kappa}$, it is most natural to select the OS~condition on the
mass of the light singlet as defining the framework for the Higgs
decays: the tree-level masses then immediately satisfy the expected
kinematical relations and only genuine hierarchies are present in the
calculation of the radiative corrections. The corresponding
predictions for the decay widths are shown as solid lines and exhibit
the expected behavior down to~$M_{a^S}\!=0$.

Radiative corrections to the decay widths into fermions are dominated
by QCD~effects (when applicable): these involve first an
ultraviolet~(UV) logarithm that we absorb within the definition of the
Yukawa couplings\,\cite{Braaten:1980yq}, then further non-logarithmic
finite corrections amounting to~\mbox{$\mathcal{O}(10$--$30\%)$} of
the tree-level widths. QED and EW/Yukawa~contributions remain at the
percent level---with the exception of~\AtoB{a^S}{c\bar{c}} in the
$R$-symmetry limit, where they reach~$15\%$ due to
the~$t_{\beta}$-suppression of the Born-level amplitude. Yet, it may
be worthwhile to underline that such mild EW/Yukawa~contributions
result from large cancellations between the vertex diagrams and the
loop corrections on the external Higgs line, so that a
separate processing of both, \EG~by defining an effective mixing
matrix---see \EG~\citere{Domingo:2017rhb}---would lead to sizable
non-predictive effects. We already emphasized this point in
\citere{Domingo:2020wiy}. In addition, the large contributions of
Higgs-mixing type do not come about as the consequence of large mixing
among the \CP-odd states---the latter remains of order~$10^{-3}$,
legitimizing the use of a non-degenerate formalism---but originate in
the much larger decay amplitude of doublet-dominated states into SM
matter. Finally, we also observe a good agreement of the fermionic
widths derived in the OS~scheme with those obtained in the original
scheme in the region where the latter produces a well-behaved
spectrum: this illustrates the weak dependence of the decay width on
the tree-level Higgs mass, which is itself related to the small size
of the EW~corrections. In fact, a much larger dependence would emerge
if we did not factor out QCD~corrections and employ a physical Higgs
mass in the corresponding piece, \IL~of the tree-level mass.

The decays into gluon and photon pairs, emerging directly at 1L order,
are much more sensitive to the employed tree-level Higgs mass, as can
be seen from the `horns' developing in the original scheme at the
limit of its applicability. This artificial sensitivity to the
tree-level mass is expected to disappear with the inclusion and
resummation of higher orders. This lengthy procedure could be bypassed
by directly upgrading the tree-level Higgs mass to its physical
value. This is possible here because no genuine weak corrections
contribute at the corresponding order, only fermion loops; in the
\CP-even case, an upgrade of the Higgs~couplings to Goldstone bosons
would also be needed under penalty of violating the Ward~identity and
generating further large spurious effects, see
\citere{Domingo:2020wiy}. Instead of following this procedure, we work
in the scheme with an OS~Higgs mass, which conveniently circumvents
such complications.

Other schemes with modified~$A_{\kappa}$ can also be employed for the
calculation of the Higgs decay widths, as long as the tree-level
spectrum does not violate the kinematical setup of the considered
process. Here, we tested the modified schemes\,A of
Sect.\,\ref{sec:CPoddmass}, with a relatively high IR-cutoff on the
tree-level Higgs mass, as well as a scheme\,C with
IR-cutoff~${\approx}\,12$\,GeV (still above
the~$b\bar{b}$~threshold). Deviations with respect to the OS~scheme
are essentially contained in the derived value of the physical
mass~$M_{a^S}$---see the comparisons in \fig{fig:PQmass} and
\fig{fig:RSmass}. Once considered as function of this mass, the
predictions of the decay widths into fermions for both modified
schemes agree with the OS~scheme at the level of~$\lsimord10\%$ of the
magnitude of the EW~contributions. Again, this illustrates the very
loose dependence of these decay channels on the chosen tree-level
mass. Nevertheless, for the diphoton and digluon final states,
where~1L is the~LO, the choice of tree-level mass (hence, of
scheme) has a critical impact on the decay widths, as we discussed
already.

\subsection{Case of a light \CP-even Higgs\label{sec:CPEdec}}

The case of a light singlet-dominated \CP-even state is much more
difficult to quantitatively describe at~1L, due to the radiative
mixing with the SM-like state. As the singlet component does not
couple to the SM decay products, the corresponding decays of the
singlet-dominated state are indeed governed by its (in
phenomenologically realistic scenarios) subdominant doublet component,
generated either at tree level or through radiative corrections. In
the \CP-odd sector, the singlet--doublet mixing was well under control
already at tree level, because of the large mass of the heavy-doublet
state, thus making the mass-splitting comparatively clean. It is not
so in the \CP-even sector, because of the simultaneous presence of two
light states with masses heavily depending on radiative corrections,
so that the mass-splitting can easily be shifted
by~$\mathcal{O}(100\%)$ by 1L (and even 2L) effects. In other words,
the tree-level description of the light states is very far from the
physical configuration and would require the inclusion of higher
orders for a precise prediction of the phenomenology. As it is, the
convergence of the perturbative series is too slow to quantitatively
assess the mixing of the singlet- and SM-like Higgs at~1L.

Let us clarify this point first. The tree-level couplings of the light
singlet to SM matter are governed by the mixing angles emerging in the
diagonalization of the mass matrix of \refeq{eq:CPevenmat}:
\begin{subequations}
\begin{align}
  U_{11} &\approx -\frac{c_{\beta}}{m^2_{h_1}\!-m^2_{h_2}}\left[
    c_{\beta}\,{\cal M}_{\text{even},13}^{2}+s_{\beta}\,{\cal M}_{\text{even},23}^{2}
    \right]
    - \frac{s_{\beta}}{m^2_{h_1}\!-m^2_{h_3}}\left[
    s_{\beta}\,{\cal M}_{\text{even},13}^{2}-c_{\beta}\,{\cal M}_{\text{even},23}^{2}
    \right],\\
  U_{12} &\approx -\frac{s_{\beta}}{m^2_{h_1}\!-m^2_{h_2}}\left[
    c_{\beta}\,{\cal M}_{\text{even},13}^{2}+s_{\beta}\,{\cal M}_{\text{even},23}^{2}
    \right]
    + \frac{c_{\beta}}{m^2_{h_1}\!-m^2_{h_3}}\left[
    s_{\beta}\,{\cal M}_{\text{even},13}^{2}-c_{\beta}\,{\cal M}_{\text{even},23}^{2}
    \right].
\end{align}
\end{subequations}
In phenomenologically realistic scenarios, considering \EG~the
constraints placed by~LEP on a light Higgs
boson\,\cite{Barate:2003sz}, these objects will typically be small, of
the order of~$\lsimord10^{-1}$. However, the Higgs spectrum is only
marginally controlled at tree level, \EG~$m^2_{h_{1,2}}$ may be
shifted by~$\gsimord100\%$ at~1L, so that~$U_{11}$ and~$U_{12}$ are
complemented by a significant loop-induced mixing (potentially of
comparable magnitude to the tree-level values). At~1L, the radiative
mixing enters the decay amplitudes via the LSZ~reduction as
\begin{align}\label{eq:decmix}
  \Amp{mix}{}{\AtoB{h_1}{\text{X}}} &= -\sum\limits_{i\neq1}
  \frac{\hat{\Sigma}^{(1)}_{1i}\big(m^2_{h_1}\big)}{m_{h_1}^2\!-m_{h_i}^2}\,
  \Amp{born}{}{\AtoB{h_i}{\text{X}}} \equiv
  \sum\limits_{i\neq1}\mathcal{Z}_{1i}\,\Amp{born}{}{\AtoB{h_i}{\text{X}}}\,,
\end{align}
where~$i\in\{2,3\}$ and~$\hat{\Sigma}^{(1)}_{1i}$ are the renormalized
off-diagonal self-energies at~1L. This expansion remains \AP~valid as
long as~$\mathcal{Z}_{1i}\ll1$. In particular, the
denominator~$m_{h_1}^2\!-m_{h_i}^2$ should be evaluated at the tree
level in this object in order to preserve the independence from
gauge-fixing parameters after combination with vertex
corrections\,\cite{Domingo:2020wiy}. We stress that~$\mathcal{Z}_{1i}$
is a scheme-, gauge- and field-dependent quantity appearing as an
intermediate step in the calculation, so that it should not be
over-interpreted. Nevertheless, its physical meaning makes it clear
that, the farther this 1L~object stands from an estimate using the
genuine physical spectrum, the larger the corrections left to higher
orders are expected to be. Here, in the case of light states, we
observe that 1L (and even~2L) effects would shift the denominator
by~$\mathcal{O}(100\%)$; this indicates that higher-order corrections
to the mixing would be numerically comparable to the 1L~version in
\refeq{eq:decmix}, which itself competes with the tree-level mixing,
as we argued above. This is but one obvious problem of the fashion in
which the Higgs mixing is assessed at~1L. In fact, the numerator
of~$\mathcal{Z}_{1i}$ is also unstable under radiative corrections,
because it is calculated with the unrealistic tree-level Higgs
couplings and mixing angles. In other words, the loop expansion
converges too slowly to claim reliable results at~1L in the
perturbative calculation performed in the SUSY~model.

One can attempt to capture the contributions of the Higgs mixing to
the decay widths in more quantitative fashions. A first strategy would
amount to resorting to the mixing formalism described in
\citeres{Domingo:2020wiy,Domingo:2021kud}.
Although~$\big\lvert\hat{\Sigma}^{(1)}_{12}\big\rvert\ll\big\lvert
m_{h_1}^2\!-m_{h_2}^2\big\rvert$ in the considered scenarios, the
closeness in mass of~$h_1$ and~$h_2$ in view of the magnitude of the
mass corrections at~1L legitimates the use of this approach: the
external momentum entering the self-energies may indeed be expanded in
the vicinity of~$m_{h_1}^2$, $m_{h_2}^2$
or~$\big(m_{h_1}^2+m_{h_2}^2\big)\big/2$ indifferently. The advantage
of this technique is that the 1L~mixing is pre-included in a
loop-corrected rotation matrix for the tree-level fields, which in
particular takes into account the large radiative mass corrections
while keeping the dependence on field counterterms and gauge-fixing
parameters to a minimum; we refer the reader to
\citeres{Domingo:2020wiy,Domingo:2021kud} for a more detailed
discussion. Nevertheless, this mixing matrix is still calculated with
the unrealistic tree-level spectrum, meaning that large higher-order
contributions are still expected to correct the off-diagonal
self-energies, \IE~the numerator of~$\mathcal{Z}_{1i}$.

\needspace{7ex} Alternatively, we can attempt to adapt the
renormalization scheme, in order to yield a tree-level Higgs spectrum
closer to the actual physical situation. We have already described
above how the mass of the light singlet could be brought~OS through a
re-definition of~$A_{\kappa}$. The status of the mass of the SM-like
state is \AP~somewhat more delicate, because, at the tree level, it
cannot be set to its physical value through a simple change of
renormalization condition on the original parameters of the
model.\footnote{This might be attempted with the renormalization
  of~$\lambda$ at very low~$t_{\beta}$, though, but this would in
  general endanger the perturbativity of this parameter.}
Nevertheless, following the suggestion of \citere{Domingo:2020wiy}
(see in particular the appendix), one may generalize the Higgs
potential of the model to an effective `non-SUSY' version, thus
allowing for an OS~renormalization of the neutral-Higgs masses. In
this approach, one introduces quartic Higgs couplings~$\ell_i$,
$i\in\{1,\cdots,7\}$, in the Higgs-doublet sector---additional terms
for the singlets could be considered as well, see
\citere{Chalons:2012qe}---together with their
counterterms~$\delta\ell_i$, and matches the corresponding model to
the SUSY~potential.\footnote{Given that this framework puts aside the
  correlation between operators inherited from~SUSY, it is \AP also
  necessary to introduce further independent counterterms in other
  sectors of the model, \EG~gauginos or sfermions, which however do
  not interest us here.} Yet, in the present paper, instead of
following the unwieldy matching conditions derived from the
identification of observables as in \citere{Domingo:2020wiy}---this
requires the introduction of a broad set of observables, hence
increasing the reliance on calculations performed in the original
description---we resort to the more convenient requirement that the
bare parameters should be unchanged in both models,
\IE~$\ell_i+\delta\ell_i=\mathcal{O}(\text{2L})$, similarly to the
condition that we imposed on~$A_{\kappa}$ at the level of the singlet
masses. Moreover, given that we specialize in the phenomenology of
light states, we need only introduce a non-vanishing~$\ell_2$ (the
coupling multiplying the new $\lvert H_u\rvert^4$~operator),
characterized by OS~conditions on the mass of the SM-like Higgs; all
the other~$\ell_i$ are kept equal to~$0$
(and~$\overline{\text{DR}}$). This choice is not unique but
convenient, as it allows for a straightforward determination of
the~$\ell_i$-s; it is also physically meaningful since the large
top/stop corrections to the SM-like mass indeed mostly project
onto~$\ell_2$. In practice we determine~$\ell_2$ by the condition
\begin{equation}
  \det{\Big[\mathcal{M}^2_{\text{even}} +
      2\,\ell_2\,v^2\,s^2_{\beta}\,\mathcal{E}^{22} -
      M^2_{h^{\text{SM}}}\,\mathds{1}\Big]}\stackrel[]{!}{=}0\,,
\end{equation}
where~$\mathcal{M}^2_{\text{even}}$ is the original (NMSSM) \CP-even
mass matrix at tree level---see
\refeq{eq:CPevenmat}---and~$\mathcal{E}^{22}$ is defined by its matrix
elements: $(\mathcal{E}^{22})_{ij}=\delta_{i2}\,\delta_{j2}$ for
$i,j\in\{1,2,3\}$. The parameter~$M^2_{h^{\text{SM}}}$ is the physical
squared Higgs mass of the SM-like state derived in the original scheme
and taking into account effects beyond~1L (\EG~$\alpha_t\,\alpha_s$,
$\alpha_t^2$ corrections, UV-resummation,
etc.\,\cite{Borowka:2014wla,Degrassi:2014pfa,Borowka:2015ura,Borowka:2018anu,Hollik:2014wea,Hollik:2014bua,Passehr:2017ufr,Goodsell:2016udb,Goodsell:2014bna,Drechsel:2016jdg,Goodsell:2014pla,Muhlleitner:2014vsa,Goodsell:2015ira,Goodsell:2015yca,Braathen:2016mmb,Biekotter:2017xmf,Stockinger:2018oxe,Bahl:2018ykj,Dao:2019qaz,Goodsell:2019zfs,Bagnaschi:2014rsa,Lee:2015uza,Vega:2015fna,Athron:2016fuq,Bagnaschi:2017xid,Bahl:2018jom,Braathen:2018htl,Bagnaschi:2019esc,Murphy:2019qpm,Bahl:2020mjy,Bahl:2020jaq,Staub:2017jnp,Harlander:2018yhj,Bahl:2019wzx,Bahl:2016brp,Bahl:2017aev,Bahl:2019hmm,Bahl:2020tuq,Harlander:2019dge,Kwasnitza:2020wli,Dao:2021khm}).
After taking into account this non-zero~$\ell_2$, the Higgs states of
the modified scheme are denoted
as~$\breve{h\,}\!_{I}=\breve{U}_{Ij}\,h^0_j$,
\mbox{$h^0_j\in\{\phi_1,\phi_2,\phi_s\}$} (the \CP-even gauge
eigenstates), together with the new tree-level diagonalization
matrix~$\breve{\mathbf{U}}$; one of the states of the modified scheme
takes the tree-level mass~$M_{h^{\text{SM}}}$. The counterterm
associated to~$\ell_2$ should satisfy~\mbox{$\delta\ell_2\approx
  -\ell_2$}. At the same time, it is constrained by the OS~definition
of the SM-like Higgs-boson
mass:~\mbox{$\breve{\Sigma}_{II}\big(M^2_{H^{\text{SM}}}\big) -
  \delta\mathcal{M}^{2\,\text{NMSSM}}_{II} -
  2\,\delta\big(\ell_2\,v^2\big)\,s^2_{\beta}\,\breve{U}_{I2}^2$} must
have a vanishing real part (up to higher-order effects),
where~$\breve{\Sigma}_{II}$ represents the diagonal self-energy
associated with the field~$\breve{h\,}\!_{I}$,
and~$\delta\mathcal{M}^{2\,\text{NMSSM}}_{II}$ the NMSSM-like
counterterm projected onto the~$\breve{h\,}\!_{I}$-direction. There is
an ambiguity at this level, because we determine~$\ell_2$ after
including several higher-order effects to the Higgs self-energies,
while we compute the decays at 1L~order. In these conditions, we
regard the choice (the superscript~$^{(1)}$ signaling restriction to
the 1L~order)
\begin{equation}\label{eq:dl2CT1}
  2\,\delta\big(\ell_2\,v^2\big)\,s^2_{\beta}\,\breve{U}_{I2}^2\stackrel[]{!}{=}
  \Real{\breve{\Sigma}^{(1)}_{II}\big(M^2_{H^{\text{SM}}}\big)
    - \delta^{(1)}\mathcal{M}^{2\,\text{NMSSM}}_{II}}
\end{equation}\enlargethispage{1.ex}%
as the most consistent one, although it violates the cancellation
of~$\ell_2+\delta\ell_2$ by the full magnitude of the effects
beyond~1L. Alternatively, the following condition, with~$I_0$ indexing
the SM-like state in the original scheme, would
satisfy~$\ell_2+\delta\ell_2\approx0$ almost exactly:
\begin{equation}\label{eq:dl2CT2}
  2\,\delta\big(\ell_2\,v^2\big)\,s^2_{\beta}\,\breve{U}_{I2}^2 \stackrel[]{!}{=}
  \Real{\breve{\Sigma}^{(1)}_{II}\big(M^2_{H^{\text{SM}}}\big)
    - \delta^{(1)}\mathcal{M}^{2\,\text{NMSSM}}_{II}}
  - \Real{\hat{\Sigma}_{I_0I_0}^{(1)}\big(m^2_{h_{I_0}}\big)}
  - 2\,\ell_2\,v^2\,s^2_{\beta}\,\breve{U}_{I2}^2\,.
\end{equation}
However, in this case, we would also need to include in the
computation of diagrams (\EG~off-diagonal Higgs self-energies or
triple-Higgs vertex) pieces that numerically mimic the impact of the
leading 2L~contributions. In Higgs decays into SM~final~states, we
find the choice between \refeq{eq:dl2CT1} or \refeq{eq:dl2CT2} to have
a numerically marginal impact. This concludes the description of this
formalism, in which we expect a better convergence of the perturbative
series of radiative corrections, because the tree-level spectrum is
closer to its physical configuration, hence resums large radiative
corrections within the tree-level mixing angles (controlling the Higgs
couplings), as well as in the new operator~$\ell_2\,\lvert
H_u\rvert^4$ encoding the correspondence between spectrum and
Higgs-to-Higgs couplings.

\begin{figure}[b!]
  \centering
  \includegraphics[width=\linewidth]{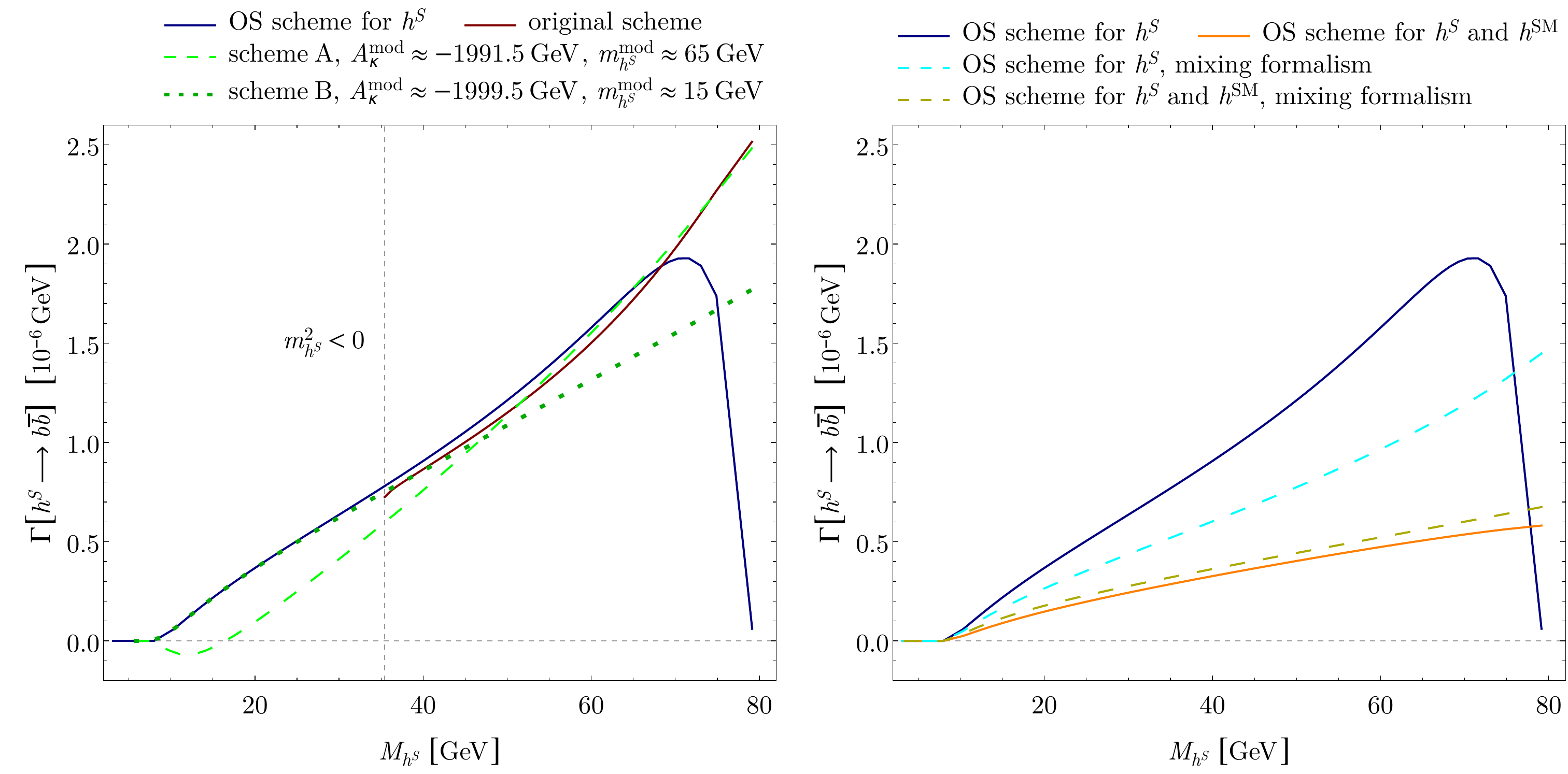}
  \caption{Partial $b\bar{b}$~decay~widths of the light \CP-even
    state~$h^S$ in the scenario\,CPE1 of \sect{sec:CPevenmass}
    with~\mbox{$\lambda=\kappa=0.05$}.\label{fig:CPEbisdec}\vspace{-2.4ex}}
\end{figure}

We now examine the situation in the scenario\,CPE1 with a small value
of~$\lambda$. The singlet--doublet mixing remains below~$10^{-2}$ at
tree level and the 1L~mixing between~$h^S = h_1$
and~\mbox{$h^{\text{SM}} = h_2$} is of comparable magnitude (though
slightly larger, depending on the scheme and point in parameter
space): the production of the light singlet at~LEP is \AP~suppressed
by about~$10^{-4}$ as compared to an SM~Higgs at the same mass. We
focus on the $b\bar{b}$~decay~channel in \fig{fig:CPEbisdec}. In the
plot on the left, all calculations are performed with the original
SUSY~Lagrangian and the expansion formalism, and only the scheme
defining~$A_{\kappa}$ is changed: OS~condition on the light singlet
mass in the blue curve, tree-level singlet mass set to a fixed value
($65$\,GeV in dashed light-green, $15$\,GeV in dotted dark-green), or
the original \DR~condition in red (failing for masses below~$35$\,GeV,
as explained in \sect{sec:CPevenmass}). All predictions provide a very
roughly comparable order of magnitude, but the dispersion already
indicates the lack of control over the singlet--doublet mixing:
occasional cancellations of the predicted decay width reflect
destructive interferences between tree-level and 1L~amplitudes, which
however do not intervene at the same point in parameter space in the
different schemes. In particular, the points
with~$m_{h^S}\!\gsim75$\,GeV lead to a near-degeneracy between~$h^S$
and~$h^{\text{SM}}$ at tree level, which, however, is very far from
the actual kinematical configuration, hence they receive a poor
description.

In the plot on the right,~$A_{\kappa}$ is always fixed by the
OS~scheme for the light-singlet mass. For reference, the solid blue
curve of the left plot is repeated here; several other approaches are
applied to attempt and control the singlet--doublet mixing. The dashed
cyan curve is obtained with the original NMSSM~Lagrangian, employing
the mixing formalism with a loop-corrected rotation matrix. The
predicted width is similar, though somewhat suppressed, to those of
the plot on the left. The other two curves employ a modified
Lagrangian including an~$\ell_2$~coupling determined by an
OS~condition for the mass of the SM-like state: the solid
orange line is derived with the strict perturbative expansion while
the dashed yellow one adds contributions in the mixing formalism. The
corresponding predictions for the decay widths are suppressed with
respect to those of the left plot. This can be understood as a
consequence of working with states that have well-separated masses
already at tree level, hence suppressing the Higgs-mixing
contributions mediated by off-diagonal self-energies. Our belief is
that mixing effects are more properly resummed in this procedure with
OS~masses. However, this debate cannot be decided at 1L~order and we
must therefore admit our incapacity to provide precise results for the
decays of the light \CP-even singlet in this scenario. All that can be
performed is to set an upper bound on the magnitude of the width. We
do not show the other decay channels: all the fermionic ones lead to a
comparable situation to that of~$b\bar{b}$; as to the bosonic
channels, the description at EW~LO can be regarded as even less
reliable.

\begin{figure}[b!]
  \centering
  \includegraphics[width=\linewidth/2]{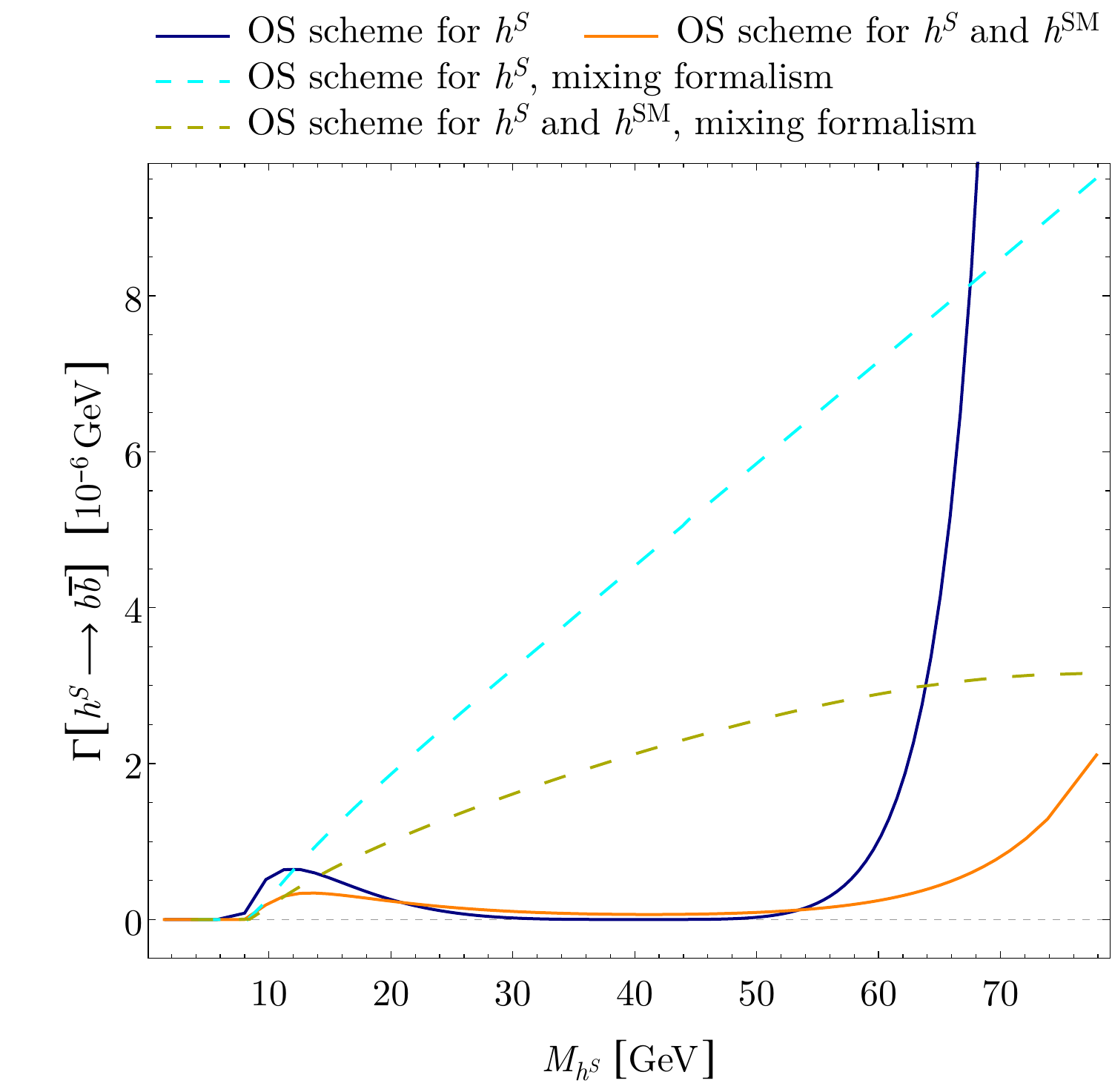}
  \caption{Partial $b\bar{b}$~decay~widths of the light \CP-even
    state~$h^S$ in the scenario\,CPE2 of Sect.\,\ref{sec:CPevenmass}
    with~\mbox{$\lambda=10\,\kappa=0.5$}.\label{fig:CPEdec}}
\end{figure}

The scenario\,CPE2 with large~$\lambda=0.5$ is considered in
\fig{fig:CPEdec}. In this case, the 1L~mixing between~$h^S = h_1$
and~$h^{\text{SM}} = h_2$ completely dominates over the tree-level one
(of order~$\simord10^{-2}$): it reaches up to~$\simord0.4$ in the
strict-expansion approach at~$m_{h^S}\!\approx75$\,GeV,
reflecting the unrealistic situation of near-degeneracy with
the~$\simord88$\,GeV SM-like Higgs at the tree level. A more realistic
assessment of the mixing via the effective mixing matrix or the
OS~scheme for the SM-like Higgs returns values
of~$\simord10^{-1}$. This still corresponds to production
cross-sections at the percent level of that of a SM~Higgs boson
at~LEP. However, given that the tree-level mixing is close to zero and
the 1L~mixing subsequently appears as the~LO, observing also that the
latter is ill-controlled, \IE~strongly dependent on the chosen scheme,
no clear predictivity can be reached at the level of the decay
widths. In fact, the dominant radiative mixing generates a negative
contribution to the width that is larger than the tree-level width;
positivity in this scenario can only be reached by also including a
1L$^2$~term, but this in turn indicates that the achieved prediction
is compatible with~$0$. The various approaches for modeling the Higgs
mixing then lead to relatively diverse results, of which only a rough
upper bound on the decay widths can be extracted. This strong
disparity is illustrated in \fig{fig:CPEdec} for the
$b\bar{b}$~channel. For the blue curve, with only~$h^S$
renormalized~OS, the mixing with the SM-like state is unrealistically
intense, especially when~$m_{h^S}\!\to m_{h^{\text{SM}}}$, \IE~at the
rightmost end of the plot. If the masses of both~$h^S$
and~$h^{\text{SM}}$ are processed~OS (orange line), the radiative
mixing remains much smaller and leads to a better-behaved
prediction. These trends persist in the mixing formalism (dashed
curves), where we should see the impact of the tree-level description
of the mixing angles on the derivation of the mixing
matrix. Nevertheless, the general shape of the dashed curves clearly
contradicts the destructive interferences between Born-level and
1L~amplitudes that are suggested by the strict expansion approach. We
must thus conclude that the predicted widths for the light singlet in
this scenario are once again dominated by the uncertainties.

\bigskip
As long as one focuses on points with suppressed singlet--doublet
mixing at tree level, the radiative mixing can be expected to be
comparatively large, since it is only protected by the EW~symmetry,
which is clearly broken at the considered scales. Then, as we showed
above, a quantitative assessment of the singlet--doublet mixing is
problematic at~1L, forbidding precise predictions for the decays of
the singlet-dominated state until higher-order corrections are fully
available. Cleaner results are likely to emerge in scenarios with a
larger singlet--doublet mixing at tree level and dominant with respect
to the radiative mixing. Such scenarios are however of lesser
phenomenological interest.

\section{Properties of the SM-like Higgs boson \label{sec:SMHiggs}}

The characteristics of the SM-like Higgs boson are of particular
interest, since experiments have been accumulating data on
corresponding observables. While we do not aim at a comprehensive
discussion of the associated phenomenology in the NMSSM here---our
benchmarks are only meant for illustration of the calculation of
radiative corrections---we can attempt to assess to which extent the
presence of new physics (light or heavy) affects the properties of the
Higgs boson at the radiative level. A detailed understanding of the
precision achieved in corresponding calculations then appears as
critical in order to avoid over-interpreting apparent deviations in
the predicted central values. We thus discuss these features in some
details below.

\subsection{SM input and SM Higgs decays\label{sec:SMinp}}

SM fermion masses and EW measurements determine a significant portion
of the parameters of the model. Here, we use comparatively recent
experimental
input\,\cite{deFlorian:2016spz,ParticleDataGroup:2020ssz}, which is
summarized in \tab{tab:SMinput}. Parameters that are given in the
minimal subtraction scheme~($\overline{\text{MS}}$) are such only from
the perspective of~QCD, and not with respect to~EW, Yukawa or
SUSY~corrections. We perform the $\overline{\text{MS}}$~running for
quark masses and the strong coupling constant to 4L~order (as well as
the $\overline{\text{MS}}$--OS~conversion), employing to that purpose
the code \texttt{RunDec}\,\cite{Chetyrkin:2000yt,Schmidt:2012az}. By
varying the input of \tab{tab:SMinput} within experimental error bands
and studying the impact of this variation on observables, one would
derive the associated parametric uncertainty: we will not discuss the
latter below, and regard the SM input as though it were `exact'.

\begin{table}[t!]
\centering\renewcommand{\arraystretch}{1.5}\vspace{3ex}
\begin{tabular}{|c | c | c | c |}
\hline
$\alpha(0)^{-1}$ & $\alpha(M_Z)^{-1}$ & $G_F$ $\big[$GeV$^{-2}\big]$ &
$\alpha^{\overline{\text{MS}}}_s(M_Z)$ \\\hline
$137.0359895$ & $128.962$ & $1.1663787\cdot 10^{-5}$ & $0.1180$ \\\hline
\end{tabular}\\[.2ex]
\begin{tabular}{|c |c | c | c | c | c | c |}
\hline
& $m^{\overline{\text{MS}}}_u(2\,\text{\small GeV})$ & $m^{\overline{\text{MS}}}_d(2\,\text{\small GeV})$ & $m^{\overline{\text{MS}}}_s(2\,\text{\small GeV})$ & $m^{\overline{\text{MS}}}_c(3\,\text{\small GeV})$ & $m^{\overline{\text{MS}}}_b(m_b)$ & $m_t$ \\\hline
[GeV] & $2.16\cdot10^{-3}$ & $4.67\cdot10^{-3}$ & $0.093$ & $0.986$ & $4.18$ & $172.76$ \\\hline\hline
& $m_e$ & $m_{\mu}$ & $m_{\tau}$ & $M_W$ & $M_Z$ & $M_H^{\text{SM}}$ \\\hline
[GeV] & $0.511\cdot10^{-3}$ & $0.105658$ & $1.7768$ & $80.379$ & $91.1876$ & $125.25$ \\\hline
\end{tabular}
\caption{SM input. Masses correspond to pole values, except when
  indicated as~$\overline{\text{MS}}$.\label{tab:SMinput}\vspace{3ex}}
\end{table}

\needspace{3ex} Our processing of the EW input is the one commonly
employed in~SM and BSM~calculations, but we detail it here for
completeness. All the EW~quantities (gauge couplings, masses of the
EW~gauge bosons, weak mixing angle, EW~v.e.v.---hence Yukawa
couplings---etc.) are defined from the following three input
quantities: the fine-structure constant~$\alpha$ and the masses of the
EW~gauge bosons~$M_{W,Z}$. The latter two are renormalized~OS at their
experimentally measured value. $\alpha$ is renormalized in the Thomson
limit, so that~$\alpha(0)$ would appear as the natural input. However,
due to the sizable dependence of the associated counterterms on
hadronic physics, it is customary to prefer either the measurement at
the
$Z$~pole,~$\alpha(M_Z)=\alpha(0)\big(1-\Delta\alpha(M_Z)\big)^{-1}$,
or the Fermi constant,
$\alpha_{G_F}\equiv\big(\sqrt{2}/\pi\big)\,G_F\,M_W^2\,\big(1-M_W^2\big/M_Z^2\big)=\alpha(0)\,(1+\Delta
r)$, as more robust input for EW~physics: the non-perturbative effects
in the electric charge counterterm are then absorbed within the
calculable quantities~$\Delta\alpha(M_Z)$ or~$\Delta r$.

The shift~$\Delta\alpha(M_Z)\approx5.9\cdot10^{-2}$ can be extracted
from data of $e^+e^-$~scattering with means of the optical
theorem\,\cite{Davier:2019can,Keshavarzi:2019abf}. The corrections to
the muon decay width,~$\Delta r$, have been computed in the
NMSSM\,\cite{Cao:2008rc,Domingo:2011uf,Stal:2015zca}. Here, we
define~\mbox{$\Delta r=\Delta r_{\text{SM}}^{\text{(4L)}}+\big(\Delta
  r_{\text{NMSSM}}^{\text{(1L)}}-\Delta
  r_{\text{SM}}^{\text{(1L)}}\big)$}; the object~\mbox{$\Delta
  r_{\text{SM}}^{\text{(4L)}}\approx3.76\cdot10^{-2}$} includes up to
leading 4L~effects calculated in the~SM and is obtained from the fit
formulae of \citeres{Awramik:2003rn,Arbuzov:2005ma}, providing a
precision of permil level on this quantity. We note in passing that
the inclusion of higher-order contributions in the~SM, \IE~the
difference~\mbox{$\Delta r_{\text{SM}}^{\text{(4L)}}-\Delta
  r_{\text{SM}}^{\text{(1L)}}\approx0.8\cdot10^{-2}$}, could be
misleading in observables computed at 1L order, since it is not clear
that SM(-like) loop corrections of higher order to such observables
would not compensate with those introduced through~$\Delta
r_{\text{SM}}^{\text{(4L)}}-\Delta r_{\text{SM}}^{\text{(1L)}}$. We
regard this as a first source of higher-order uncertainty on the
EW~corrections. Coming back to the calculation of~$\Delta r$, a few
classes of higher-order BSM~contributions to~$\Delta r$ in SUSY~models
have been computed in
\citeres{Djouadi:1996pa,Djouadi:1998sq,Heinemeyer:2002jq,Haestier:2005ja}
and exploited in the context of the NMSSM
before\,\cite{Domingo:2011uf,Stal:2015zca}. Nevertheless, we restrict
ourselves to strict 1L contributions from new physics here. The latter
typically contribute to~$\Delta r$ at the level
of~$\mathcal{O}(10^{-3})$ in the considered scenarios.

In practice, we present only results obtained with the
parameter~$\alpha_{G_F}$ as defined above in this paper. Nevertheless,
also performing the calculation in the scheme associated
with~$\alpha(M_Z)$ provides a useful alternative in order to assess
the uncertainty associated with the choice
of~$\alpha$. As~$\alpha_{G_F}$ and~$\alpha(M_Z)$ differ
by~$\delta\alpha/\alpha\sim2.2\%$, the calculation \AP~cannot claim a
precision narrower than this value on the last controlled EW~order. We
count this uncertainty, dominated by effects of SM-type, as being of
the parametric class.

\needspace{10ex} Before considering new-physics effects to the Higgs
decays, it seems desirable to analyze the situation in the SM
beforehand, and evaluate corresponding decay widths at exactly the
same order as later in the BSM model to serve as a reference. We
consider the following channels (for an SM Higgs boson~$H^{\text{SM}}$
with OS~mass~$M_{H^{\mathrm{SM}}}$):
\begin{itemize}
\item Decays into quarks ($b\bar{b}$, $c\bar{c}$) and leptons
  ($\tau^+\tau^-$): we evaluate these channels at the same order as
  commonly done in the SM\,\cite{deFlorian:2016spz}, \IE~we include
  QCD~corrections to 4L~order---see
  \EG~\citeres{Chetyrkin:1995pd,Baikov:2005rw} or the summary in
  \citere{Spira:2016ztx}, taking also finite-fermion-mass effects into
  account at 1L\,\cite{Braaten:1980yq,Drees:1990dq}---as well as~QED
  and EW~corrections at~1L. The corresponding theoretical uncertainty
  has been evaluated to below~$1\%$ in \citere{deFlorian:2016spz} at
  the level of the branching ratios, which are expected to be largely
  insensitive to the choice of input for~$\alpha$. Our results for the
  fermionic Higgs decay widths employing~$\alpha_{G_F}$ are consistent
  with those published by the
  LHC~Cross-section~Working~Group\,\cite{deFlorian:2016spz,LHCXSWGBR-www},
  obtained with
  \texttt{HDECAY}\,\cite{Djouadi:1997yw,Djouadi:2018xqq}, up to a few
  permil. However, we stress that the choice of~$\alpha(M_Z)$ for
  input affects the decay widths by~$1$--$2\%$, to which further
  parametric uncertainty should be added, from \EG~$\alpha_s$ or the
  fermion masses---the latter typically amount to the percent
  level\,\cite{LHCXSWGBR-www}.
\item Decays into EW~gauge bosons ($WW$, $ZZ$): for an SM~Higgs boson,
  these decays proceed off-shell and in fact approximate the more
  complete decay channels with four fermions in the final state. Given
  that a full 1L~analysis in the~NMSSM is beyond our current means, we
  employ a tree-level description---see \EG~Eqs.\,(37--39) in
  \citere{Spira:2016ztx}. The standard in the~SM is given by the
  calculation of
  \texttt{Prophecy4f}\,\cite{Bredenstein:2006rh,Bredenstein:2006nk,Bredenstein:2006ha},
  with an estimated theoretical uncertainty below
  percent\,\cite{deFlorian:2016spz}. The corrections to the tree-level
  results are of order~$5\%$. In our NMSSM~scenarios, we
  rescale the results for the SM-like Higgs
  accordingly, which is a way of putting SM-like effects under
  control.
\item Radiative decays into gauge bosons ($\gamma\gamma$, $gg$): we
  evaluate these decay widths at leading EW~order, including
  $\mathcal{O}(\alpha_s)$~corrections to the diphoton
  channel\,\cite{Spira:1995rr} and up to
  $\mathcal{O}\big(\alpha_s^5\big)$~corrections to the digluon
  channel\,\cite{Baikov:2006ch}---the
  $\mathcal{O}\big(\alpha_s^3\big)$~SUSY~corrections\,\cite{Muhlleitner:2006wx}
  will be included in the~NMSSM as well. We directly consider the
  digluon~width with radiation of five light-quark flavors, even
  though at least bottom (and possibly charm) quarks should lead to a
  distinguishable final state. In these conditions, the uncertainty
  from EW~higher orders is dominant, amounting to~$\mathcal{O}(20\%)$
  for the diphoton~channel. Comparison with \texttt{HDECAY}, where
  2L~EW~corrections are included for the SM~Higgs decays, reveals that
  the leading EW~order for the diphoton~channel has an improved
  predictivity after rescaling of the decay width by a
  factor~$\big(\alpha(0)/\alpha_{G_F}\big)^2$; we apply this
  correction. Our prediction in the digluon~channel
  remains~$\simord6\%$ off. Beyond its impact on the total width, this
  latter decay channel is mostly relevant as an estimate of the
  effective Higgs coupling to gluons, which is relevant for Higgs
  production at hadron colliders.
\end{itemize}
We present our results for the decays of a SM~Higgs boson with a mass
of~$125.25$\,GeV in the second column of \tab{tab:SMprop}. We stress
that they are not meant as state-of-the-art evaluations of these Higgs
properties in the~SM, but only as reference for comparison with
BSM~results that are calculated at a similar order in the loop
expansion. There, we consider the decay widths rather than the
branching ratios, the latter being more useful for comparison with
experiment. The reason for this choice is that the channels known with
least precision in our calculation (bosonic channels) contaminate the
total width, hence increase the theoretical uncertainty for the more
precisely known (fermionic) channels. In addition, non-standard
channels also contribute to the total width in the~NMSSM. Thus, a
comparison with the~SM is easier at the level of the decay widths.

\subsection{Mass and decays of the SM-like Higgs boson into SM particles}

Now we turn to the SM-like Higgs boson of the considered
NMSSM~scenarios. A first observable that has received considerable
attention is the mass of this particular
state\,\cite{Slavich:2020zjv}. Radiative corrections of Yukawa-type
have a sizable impact on these predictions. We are in a position to
include corresponding effects at the 2L~level and resum UV logarithms
for heavy-squark~spectra according to the description in
\citere{Domingo:2021kud}. EW~and NMSSM-specific effects are included
at~1L. A large value of~$\lambda$ generates a contribution already at
tree level, which is mostly relevant for moderate~$t_{\beta}$; in the
considered benchmarks, this amounts to an effect of at most a few~GeV
in the PQ-scenario with~$t_{\beta}=5$ and~$\simord20$\,GeV in the
scenario\,CPE2 with~$t_{\beta}=2$. On the other hand, we do not
observe any significant NMSSM-specific effect at~1L in the considered
scenarios. The properties of the squark~sector offer an obvious handle
on the SM-like Higgs mass observable, almost independently from the
characteristics of the singlet~Higgs, and we chose soft masses and
trilinear couplings so as to accommodate a central value of the mass
prediction compatible with~$125.25$\,GeV. Higher-order uncertainties
are dominated by EW~logarithms at~2L, large squark-mixing effects, as
well as NMSSM-specific contributions of 2L~order. All these factors
accumulate to a higher-order uncertainty at the
GeV~level\,\cite{Staub:2015aea,Drechsel:2016htw,Bahl:2019hmm}.
Furthermore, alternating the schemes for the input of~$\alpha$
between~$\alpha_{G_F}$ and~$\alpha(M_Z)$ results in variations
of~$\mathcal{O}(1\%)$, dominated by the fluctuations of the
uncorrected $\alpha_q\,\alpha_s$~contributions. In turn, these
theoretical and parametric uncertainties on the mass determination can
be regarded as a parametric uncertainty at the level of the Higgs
decays (which we ignore below).

\bigskip
The SM-like decay channels listed in \sect{sec:SMinp} can be
considered to the end of testing the emergence of BSM~physics in Higgs
interactions. New physics intervenes here in two different fashions:
\begin{enumerate}
\item SUSY and THDM effects decouple in the limit of
  heavy~$M_{H^{\pm}}$ and~$M_{\text{SUSY}}$. The EW~symmetry protects
  the relation between the mass of the SM~fermions or gauge bosons and
  the couplings of the SM-Higgs component (\IE~the exact
  $SU(2)_{\text{L}}$~partner of the Goldstone~bosons) to these same
  particles; radiative corrections projecting onto that direction are
  thus `invisible' to a large extent. Contributions to orthogonal
  directions are suppressed as~$v^2\big/M_{H^{\pm}}^2$, where only the
  THDM~sector causes deviations at tree level. The SUSY~sector
  intervenes at the loop~level, hence generates contributions that are
  suppressed with an additional loop~factor, although partially
  compensated by a UV-logarithm. In the scenarios under consideration,
  with~$M_{H^{\pm}}$ and~$M_{\text{SUSY}}$ in the TeV~range, the
  contributions from heavy new~physics are thus expected to remain at
  the percent~level, \IE~they are at most comparable to the
  theoretical uncertainty achieved in the calculation of the decay
  widths.
\item Light-singlet states may affect the phenomenology of the SM-like
  Higgs in several fashions. First, they open new decay~channels,
  which we discuss in the next subsection. Then, mixing of the SM-like
  Higgs with a singlet-dominated state leads to an exchange of
  properties between the two states: with \CP-conservation, this
  effect concerns only scenarios with a light \CP-even singlet. We
  have discussed in \sect{sec:CPEdec} how the lack of control on this
  mixing at~1L forbade a quantitative determination of the
  decay~widths for the singlet-dominated state. The situation of the
  SM-like Higgs is less critical, since the Born-level amplitude is
  expected to be dominant. However, the accuracy of the predicted
  decay~widths is bound to suffer from the unclear status of the
  mixing.
  These are the essential effects that we aim at assessing in the
  considered scenarios.
\end{enumerate}

\needspace{5ex} The evaluation of the decay widths follows the
description given in \sect{sec:SMinp} at the level of the SM-like
Higgs boson, \IE~includes complete 1L~contributions (now also
involving the~SUSY and extended-Higgs~sectors)---except for the~$WW$
and $ZZ$~channels, which remain at tree level---as well as
higher-order QCD~corrections to the same available order as in
the~SM. However, several new features need to be considered when
interpreting the results and estimating the uncertainty budget.
\begin{itemize}
\item In order to avoid introducing artificial symmetry-violating
  effects and explicit dependence of the observable on the choice of
  field renormalization, one should evaluate the loop functions in the
  expansion formalism described in
  \EG~\citeres{Domingo:2020wiy,Domingo:2021kud}. In other words, the
  external momenta entering the loop functions need always be expanded
  in the vicinity of the tree-level spectrum.
\item As we observed before, the mass of the SM-like Higgs boson of
  the SUSY~model cannot be renormalized~OS using only the original
  parameters of the SUSY~Lagrangian, in contrast to the situation in
  the~SM. As is well-known, the tree-level mass is typically
  below~$M_Z$, even though NMSSM-specific effects can circumvent this
  bound at large~$\lambda$ and low~$t_{\beta}\lsim2$. This means that
  the tree-level spectrum is always significantly away from the actual
  kinematical configuration, and that 1L (and even~2L) mass-shifts can
  be comparable to the tree-level mass. This is of limited consequence
  for processes where EW~effects are controlled at~NLO, \IE~the Higgs
  decays into SM~fermions in our description. However, we have already
  seen in \fig{fig:A1dec} that this feature is problematic for the
  diphoton and digluon~channels, where only the~EW~LO is known. This
  difficulty can be circumvented by upgrading the tree-level Higgs
  mass to its physical value in all its occurrences, \IE~in external
  momenta as well as in the Higgs potential used in the contributing
  loops. The latter can be achieved straightforwardly in the digluon
  and diphoton case as only the Higgs-couplings to charged Goldstone
  bosons need an upgrade
  then\,\cite{Benbrik:2012rm,Domingo:2018uim}---it is much more
  involved in the general case\,\cite{Domingo:2020wiy}. Here, we
  resort to the `cleaner'~method described in \sect{sec:CPEdec},
  introducing new parameters in the Higgs potential that allow for an
  OS~renormalization of the SM-like Higgs mass. For all decay
  channels, comparison of the widths derived with the original
  spectrum and those obtained with an OS~SM-like Higgs allows to
  estimate the impact of the unrealistic tree-level spectrum on the
  accuracy of the predictions.
\item We must also account for the limited control over the
  Higgs~mixing at~1L in scenarios containing a light singlet-dominated
  state~$h^S = h_1$, as explained in \sect{sec:CPEdec}. Once again, we
  can consider several approaches in order to assess the impact of
  these mixing effects at the level of the decays of the SM-like
  Higgs~$h^{\text{SM}} = h_2$: first, given
  that~$\big\lvert\hat{\Sigma}^{(1)}_{21}\big\rvert\ll\big\lvert
  m_{h_2}^2\!-m_{h_1}^2\big\rvert$ in the considered scenarios, the
  strict perturbative expansion generally falls in its regime of
  validity; then, it is also possible to resort to the
  mixing~formalism of \citeres{Domingo:2020wiy,Domingo:2021kud}, which
  is legitimate in view of the comparable magnitude of tree-level and
  1L~contributions to the mass-splitting; finally, we can redefine the
  Higgs~potential so as to allow for an OS~definition of the relevant
  Higgs masses, as explained above. We reserve our opinion as to which
  method provides the most reliable results, a question that can only
  be decided through the inclusion of higher-order contributions to
  the decay~widths.
\item In the case of Higgs decays into EW~gauge bosons, where the
  actual calculation is at tree level, the rescaling by the more
  complete SM~result of~\texttt{Prophecy4f} remains somewhat
  meaningful in that non-SM-Higgs~components do not couple to~$W$-s
  and~$Z$-s at tree level, so that corresponding BSM~contributions (by
  new particles with mass~$M_{\text{BSM}}$) would be suppressed by
  both a factor~$v^2\big/M_{\text{BSM}}^2$ and a
  loop~factor. Nevertheless, the difficulty of quantitatively
  assessing the singlet--doublet mixing in the presence of a light
  \CP-even singlet limits the reliability of such a tree-level
  estimate in corresponding scenarios. In addition, we stress that the
  light singlets open up new contributions to the four-fermion decays:
  these are counted separately and discussed in the following
  subsection.
\end{itemize}
For all these reasons, it is clear that the precision achieved in the
calculation of the decay widths is necessarily inferior to that
associated with the SM at the same order. A reduction of the
theoretical uncertainties below percent level would thus require the
inclusion of higher orders. In the presence of a light \CP-even
singlet mixing with the SM-like state, the higher-order uncertainty
may be even significantly larger than percent, due to the lack of
control on radiative mixing.

\begin{table}[p!]
  \tikzset{%
    diagonal fill/.style 2 args={%
        double color fill={#1}{#2},
        shading angle=45,
        opacity=0.4},
    stripes/.style 2 args={%
        shade,
        stripes fill={#1}{#2},
        shading angle=45,
        opacity=0.4}
   }
  \renewcommand{\arraystretch}{1.5}
\begin{tabular}{||r@{\,}l || c || c | c || c | c | c | c | c || c}
\hhline{|t:==:t:=:t:==:t:=====:t:=}
\multicolumn{2}{||c||}{Scenario} & SM & \multicolumn{2}{c||}{$R$-symmetric} &
\multicolumn{5}{c||}{PQ-limit} & \\\hhline{|:==::=::==::=====::=}
$A_{\kappa}^{\text{orig}}$ & [GeV] & --- & \multicolumn{2}{c||}{$-1.5$} &
\multicolumn{5}{c||}{$-40$} & \\\hhline{||--||-||--||-----||-}
\cellBG{diagonal fill={Dandelion}{Dandelion}}{2}{||@{}}{c}{||}{$A_{\kappa}$ scheme} &
--- & \multicolumn{2}{c||}{original} &
\multicolumn{5}{c||}{original} & \\\hhline{||--||-||--||-----||-}
\cellBG{diagonal fill={YellowGreen}{YellowGreen}}{2}{||@{}}{c}{||}{$\ell_2$ scheme} &
--- &
$0$ &
\cellBG{diagonal fill={YellowGreen}{YellowGreen}}{1}{}{c}{||}{$h^{\text{SM}}$ OS} &
$0$ &
\cellBG{diagonal fill={YellowGreen}{YellowGreen}}{2}{}{c}{|}{$h^{\text{SM}}$ OS} &
$0$ &
\cellBG{diagonal fill={YellowGreen}{YellowGreen}}{1}{}{c}{||}{$h^{\text{SM}}$ OS} &
\\\hhline{||--||-||--||-----||-}
\cellBG{diagonal fill={SkyBlue}{SkyBlue}}{2}{||@{}}{c}{||}{$\hat{\kappa}^2$ scheme} &
--- &
\multicolumn{2}{c||}{$0$} &
\multicolumn{2}{c|}{$0$} &
\cellBG{diagonal fill={SkyBlue}{SkyBlue}}{1}{}{c}{|}{$h^S$ OS} &
$0$ &\\\hhline{||--||-||--||-----||-}
\multicolumn{2}{||c||}{Mixing formalism} & --- & \multicolumn{2}{c||}{No} & \multicolumn{3}{c|}{No} & \multicolumn{2}{c||}{\textbf{Yes}} & \\\hhline{|:==::=::==::=====::=}
$\Gamma_{b\bar{b}}$ & [MeV] & $2.377$ &
$2.419$ &
\cellBG{stripes={YellowGreen}{YellowGreen}}{1}{}{c}{||}{$2.405$} &
$2.362$ &
\cellBG{stripes={YellowGreen}{YellowGreen}}{1}{}{c}{|}{$2.407$} &
\cellBG{stripes={YellowGreen}{SkyBlue}}{1}{}{c}{|}{$2.428$} &
$\mathbf{2.378}$ &
\cellBG{stripes={YellowGreen}{YellowGreen}}{1}{}{c}{||}{$\mathbf{2.420}$} & \\\hhline{||--||-||--||-----||-}
$\Gamma_{\tau^+\tau^-}$ & [MeV] & $0.257$ &
$0.262$ &
\cellBG{stripes={YellowGreen}{YellowGreen}}{1}{}{c}{||}{$0.261$} &
$0.256$ &
\cellBG{stripes={YellowGreen}{YellowGreen}}{1}{}{c}{|}{$0.262$} &
\cellBG{stripes={YellowGreen}{SkyBlue}}{1}{}{c}{|}{$0.264$} &
$\mathbf{0.258}$ &
\cellBG{stripes={YellowGreen}{YellowGreen}}{1}{}{c}{||}{$\mathbf{0.263}$} & \\\hhline{||--||-||--||-----||-}
$\Gamma_{c\bar{c}}$ & [MeV] & $0.119$ &
$0.118$ &
\cellBG{stripes={YellowGreen}{YellowGreen}}{1}{}{c}{||}{$0.118$} &
$0.118$ &
\cellBG{stripes={YellowGreen}{YellowGreen}}{1}{}{c}{|}{$0.118$} &
\cellBG{stripes={YellowGreen}{SkyBlue}}{1}{}{c}{|}{$0.118$} &
$\mathbf{0.118}$ &
\cellBG{stripes={YellowGreen}{YellowGreen}}{1}{}{c}{||}{$\mathbf{0.117}$} & \\\hhline{||--||-||--||-----||-}
$\Gamma_{W^+W^-}$ & [MeV] & $0.898$ &
$0.898$ &
\cellBG{stripes={YellowGreen}{YellowGreen}}{1}{}{c}{||}{$0.898$} &
$0.898$ &
\cellBG{stripes={YellowGreen}{YellowGreen}}{1}{}{c}{|}{$0.898$} &
\cellBG{stripes={YellowGreen}{SkyBlue}}{1}{}{c}{|}{$0.898$} &
$\mathbf{0.897}$ &
\cellBG{stripes={YellowGreen}{YellowGreen}}{1}{}{c}{||}{$\mathbf{0.893}$} & \\\hhline{||--||-||--||-----||-}
$\Gamma_{ZZ}$ & [MeV] & $0.110$ &
$0.110$ &
\cellBG{stripes={YellowGreen}{YellowGreen}}{1}{}{c}{||}{$0.110$} &
$0.110$ &
\cellBG{stripes={YellowGreen}{YellowGreen}}{1}{}{c}{|}{$0.110$} &
\cellBG{stripes={YellowGreen}{SkyBlue}}{1}{}{c}{|}{$0.110$} &
$\mathbf{0.110}$ &
\cellBG{stripes={YellowGreen}{YellowGreen}}{1}{}{c}{||}{$\mathbf{0.110}$} & \\\hhline{||--||-||--||-----||-}
$\Gamma_{gg}$ & [MeV] & $0.316$ &
$0.290$ &
\cellBG{stripes={YellowGreen}{YellowGreen}}{1}{}{c}{||}{$0.314$} &
$0.290$ &
\cellBG{stripes={YellowGreen}{YellowGreen}}{1}{}{c}{|}{$0.313$} &
\cellBG{stripes={YellowGreen}{SkyBlue}}{1}{}{c}{|}{$0.313$} &
$\mathbf{0.290}$ &
\cellBG{stripes={YellowGreen}{YellowGreen}}{1}{}{c}{||}{$\mathbf{0.312}$} & \\\hhline{||--||-||--||-----||-}
$\Gamma_{\gamma\gamma}$ & $\big[10^{-3}$\,MeV$\big]$ & $9.5$ &
$7.6$ &
\cellBG{stripes={YellowGreen}{YellowGreen}}{1}{}{c}{||}{$9.6$} &
$7.6$ &
\cellBG{stripes={YellowGreen}{YellowGreen}}{1}{}{c}{|}{$9.6$} &
\cellBG{stripes={YellowGreen}{SkyBlue}}{1}{}{c}{|}{$9.6$} &
$\mathbf{7.6}$ &
\cellBG{stripes={YellowGreen}{YellowGreen}}{1}{}{c}{||}{$\mathbf{9.6}$} & \\\hhline{|b:==:b:=:b:==:b:=====:b:=}
\end{tabular}\vspace{0.3cm}
\newline\null\hfill\begin{tabular}{ c || c | c | c | c | c | c || c | c | c | c ||}
\hhline{=:t:======:t:====:t|}
& \multicolumn{6}{c||}{CPE1: $\lambda=\kappa=0.05$, $t_{\beta}=10$} &
\multicolumn{4}{c||}{CPE2: $\lambda=10\,\kappa=0.5$, $t_{\beta}=2$}\\\hhline{=::======::====:|}
& \multicolumn{6}{c||}{$-1994.1$} & \multicolumn{4}{c||}{$-438.4$}\\\hhline{-||------||----||}
&
\multicolumn{2}{c|}{original} &
\cellBG{diagonal fill={Dandelion}{Dandelion}}{2}{}{c}{|}{$h^S$ OS} &
\multicolumn{2}{c||}{original} &
\cellBGend{diagonal fill={Dandelion}{Dandelion}}{4}{}{c}{||}{$h^S$ OS}\\\hhline{-||------||----||}
&
$0$ &
\cellBG{diagonal fill={YellowGreen}{YellowGreen}}{1}{}{c}{|}{$h^{\text{SM}}$ OS} &
$0$ &
\cellBG{diagonal fill={YellowGreen}{YellowGreen}}{1}{}{c}{|}{$h^{\text{SM}}$ OS} &
$0$ &
\cellBG{diagonal fill={YellowGreen}{YellowGreen}}{1}{}{c}{||}{$h^{\text{SM}}$ OS} &
$0$ &
\cellBG{diagonal fill={YellowGreen}{YellowGreen}}{1}{}{c}{|}{$h^{\text{SM}}$ OS} &
$0$ &
\cellBGend{diagonal fill={YellowGreen}{YellowGreen}}{1}{}{c}{||}{$h^{\text{SM}}$ OS}\\\hhline{-||------||----||}
& \multicolumn{6}{c||}{$0$} & \multicolumn{4}{c||}{$0$} \\\hhline{-||------||----||}
& \multicolumn{4}{c|}{No} & \multicolumn{2}{c||}{\textbf{Yes}} & \multicolumn{2}{c|}{No} & \multicolumn{2}{c||}{\textbf{Yes}} \\\hhline{=::======::====:|}
&
$2.390$ &
\cellBG{stripes={YellowGreen}{YellowGreen}}{1}{}{c}{|}{$2.383$} &
\cellBG{stripes={Dandelion}{Dandelion}}{1}{}{c}{|}{$2.390$} &
\cellBG{stripes={Dandelion}{YellowGreen}}{1}{}{c}{|}{$2.383$} &
$\mathbf{2.391}$ &
\cellBG{stripes={YellowGreen}{YellowGreen}}{1}{}{c}{||}{$\mathbf{2.383}$} &
\cellBG{stripes={Dandelion}{Dandelion}}{1}{}{c}{|}{\,$2.441$\,} &
\cellBG{stripes={Dandelion}{YellowGreen}}{1}{}{c}{|}{\,$2.448$\,} &
\cellBG{stripes={Dandelion}{Dandelion}}{1}{}{c}{|}{$\mathbf{2.426}$} &
\cellBGend{stripes={Dandelion}{YellowGreen}}{1}{}{c}{||}{$\mathbf{2.426}$}\\\hhline{-||------||----||}
&
$0.259$ &
\cellBG{stripes={YellowGreen}{YellowGreen}}{1}{}{c}{|}{$0.259$} &
\cellBG{stripes={Dandelion}{Dandelion}}{1}{}{c}{|}{$0.259$} &
\cellBG{stripes={Dandelion}{YellowGreen}}{1}{}{c}{|}{$0.259$} &
$\mathbf{0.259}$ &
\cellBG{stripes={YellowGreen}{YellowGreen}}{1}{}{c}{||}{$\mathbf{0.259}$} &
\cellBG{stripes={Dandelion}{Dandelion}}{1}{}{c}{|}{$0.265$} &
\cellBG{stripes={Dandelion}{YellowGreen}}{1}{}{c}{|}{$0.266$} &
\cellBG{stripes={Dandelion}{Dandelion}}{1}{}{c}{|}{$\mathbf{0.263}$} &
\cellBGend{stripes={Dandelion}{YellowGreen}}{1}{}{c}{||}{$\mathbf{0.264}$}\\\hhline{-||------||----||}
&
$0.118$ &
\cellBG{stripes={YellowGreen}{YellowGreen}}{1}{}{c}{|}{$0.118$} &
\cellBG{stripes={Dandelion}{Dandelion}}{1}{}{c}{|}{$0.118$} &
\cellBG{stripes={Dandelion}{YellowGreen}}{1}{}{c}{|}{$0.118$} &
$\mathbf{0.118}$ &
\cellBG{stripes={YellowGreen}{YellowGreen}}{1}{}{c}{||}{$\mathbf{0.118}$} &
\cellBG{stripes={Dandelion}{Dandelion}}{1}{}{c}{|}{$0.118$} &
\cellBG{stripes={Dandelion}{YellowGreen}}{1}{}{c}{|}{$0.118$} &
\cellBG{stripes={Dandelion}{Dandelion}}{1}{}{c}{|}{$\mathbf{0.118}$} &
\cellBGend{stripes={Dandelion}{YellowGreen}}{1}{}{c}{||}{$\mathbf{0.117}$}\\\hhline{-||------||----||}
&
$0.898$ &
\cellBG{stripes={YellowGreen}{YellowGreen}}{1}{}{c}{|}{$0.898$} &
\cellBG{stripes={Dandelion}{Dandelion}}{1}{}{c}{|}{$0.898$} &
\cellBG{stripes={Dandelion}{YellowGreen}}{1}{}{c}{|}{$0.898$} &
$\mathbf{0.898}$ &
\cellBG{stripes={YellowGreen}{YellowGreen}}{1}{}{c}{||}{$\mathbf{0.898}$} &
\cellBG{stripes={Dandelion}{Dandelion}}{1}{}{c}{|}{$0.897$} &
\cellBG{stripes={Dandelion}{YellowGreen}}{1}{}{c}{|}{$0.897$} &
\cellBG{stripes={Dandelion}{Dandelion}}{1}{}{c}{|}{$\mathbf{0.898}$} &
\cellBGend{stripes={Dandelion}{YellowGreen}}{1}{}{c}{||}{$\mathbf{0.892}$}\\\hhline{-||------||----||}
&
$0.110$ &
\cellBG{stripes={YellowGreen}{YellowGreen}}{1}{}{c}{|}{$0.110$} &
\cellBG{stripes={Dandelion}{Dandelion}}{1}{}{c}{|}{$0.110$} &
\cellBG{stripes={Dandelion}{YellowGreen}}{1}{}{c}{|}{$0.110$} &
$\mathbf{0.110}$ &
\cellBG{stripes={YellowGreen}{YellowGreen}}{1}{}{c}{||}{$\mathbf{0.110}$} &
\cellBG{stripes={Dandelion}{Dandelion}}{1}{}{c}{|}{$0.110$} &
\cellBG{stripes={Dandelion}{YellowGreen}}{1}{}{c}{|}{$0.110$} &
\cellBG{stripes={Dandelion}{Dandelion}}{1}{}{c}{|}{$\mathbf{0.110}$} &
\cellBGend{stripes={Dandelion}{YellowGreen}}{1}{}{c}{||}{$\mathbf{0.110}$}\\\hhline{-||------||----||}
&
$0.291$ &
\cellBG{stripes={YellowGreen}{YellowGreen}}{1}{}{c}{|}{$0.314$} &
\cellBG{stripes={Dandelion}{Dandelion}}{1}{}{c}{|}{$0.291$} &
\cellBG{stripes={Dandelion}{YellowGreen}}{1}{}{c}{|}{$0.314$} &
$\mathbf{0.291}$ &
\cellBG{stripes={YellowGreen}{YellowGreen}}{1}{}{c}{||}{$\mathbf{0.314}$} &
\cellBG{stripes={Dandelion}{Dandelion}}{1}{}{c}{|}{$0.293$} &
\cellBG{stripes={Dandelion}{YellowGreen}}{1}{}{c}{|}{$0.316$} &
\cellBG{stripes={Dandelion}{Dandelion}}{1}{}{c}{|}{$\mathbf{0.293}$} &
\cellBGend{stripes={Dandelion}{YellowGreen}}{1}{}{c}{||}{$\mathbf{0.313}$}\\\hhline{-||------||----||}
&
$0.291$ &
\cellBG{stripes={YellowGreen}{YellowGreen}}{1}{}{c}{|}{$0.314$} &
\cellBG{stripes={Dandelion}{Dandelion}}{1}{}{c}{|}{$0.291$} &
\cellBG{stripes={Dandelion}{YellowGreen}}{1}{}{c}{|}{$0.314$} &
$\mathbf{0.291}$ &
\cellBG{stripes={YellowGreen}{YellowGreen}}{1}{}{c}{||}{$\mathbf{0.314}$} &
\cellBG{stripes={Dandelion}{Dandelion}}{1}{}{c}{|}{$0.293$} &
\cellBG{stripes={Dandelion}{YellowGreen}}{1}{}{c}{|}{$0.316$} &
\cellBG{stripes={Dandelion}{Dandelion}}{1}{}{c}{|}{$\mathbf{0.293}$} &
\cellBGend{stripes={Dandelion}{YellowGreen}}{1}{}{c}{||}{$\mathbf{0.313}$}\\\hhline{-||------||----||}
&
$7.6$ &
\cellBG{stripes={YellowGreen}{YellowGreen}}{1}{}{c}{|}{$9.6$} &
\cellBG{stripes={Dandelion}{Dandelion}}{1}{}{c}{|}{$7.6$} &
\cellBG{stripes={Dandelion}{YellowGreen}}{1}{}{c}{|}{$9.6$} &
$\mathbf{7.6}$ &
\cellBG{stripes={YellowGreen}{YellowGreen}}{1}{}{c}{||}{$\mathbf{9.6}$} &
\cellBG{stripes={Dandelion}{Dandelion}}{1}{}{c}{|}{$7.6$} &
\cellBG{stripes={Dandelion}{YellowGreen}}{1}{}{c}{|}{$9.6$} &
\cellBG{stripes={Dandelion}{Dandelion}}{1}{}{c}{|}{$\mathbf{7.6}$} &
\cellBGend{stripes={Dandelion}{YellowGreen}}{1}{}{c}{||}{$\mathbf{9.6}$}\\\hhline{=:b:======:b:====:b|}
\end{tabular}
\caption{Decay widths of the SM-like Higgs candidate in the standard
  channels, for the SM (second column) and the scenarios of
  \sect{sec:mass} and \sect{sec:singletdec}. The mass of the SM-like
  Higgs is $125.25$\,GeV. The five first rows provide details on the
  chosen input and scheme: value of $A_{\kappa}$ in the original
  scheme, chosen renormalization of $A_{\kappa}$, $\ell_2$ and
  $\hat{\kappa}^2$ for the calculation (see text for the definition of
  these couplings), use of a mixing formalism for the system of the
  \CP-even singlet-like and SM-like Higgs bosons. \label{tab:SMprop}}
\end{table}

Predictions for the decay widths into SM-particles (we do not consider
light fermions or $Z\gamma$) are collected in \tab{tab:SMprop} for the
considered scenarios. As these properties hardly depend on the
variation of~$A_{\kappa}$ that we performed in the earlier sections
(at least as long as the tree-level masses of~$h_1$ and~$h_2$ are not
too close), we only show results for the
value~$A_{\kappa}\stackrel[]{!}{=}-1.5$\,GeV in the
$R$-symmetry-inspired scenario, $-40$\,GeV in the PQ-inspired
scenario, $-1994$\,GeV in the scenario\,CPE1
with~$\lambda=\kappa=0.05$, $-438$\,GeV in the scenario\,CPE2
with~$\lambda=10\,\kappa=0.5$. We observe that all the decay widths
are consistent with an SM~interpretation (see second column) within a
few~percent (\IE~a magnitude comparable to the theoretical
uncertainty).

For the scenario in the $R$-symmetric limit, we perform the
calculation of the decay widths both with the original SUSY~Lagrangian
and in the version with extended Higgs potential and OS~SM-like Higgs
(\IE~$\ell_2\neq0$): numerical differences are minor except for the
diphoton and digluon channels, where the version with extended Higgs
potential is \AP~more trustworthy. For the fermionic decays, we
observe that the sizable difference between tree-level and physical
masses for the SM-like Higgs~$h^{\text{SM}} = h_1$ costs somewhat more
than~$0.5\%$ on the accuracy of the prediction. Now turning to the
phenomenology, the comparison of the fermionic decay widths seems to
indicate that the SM-like Higgs of this scenario has a slightly
increased affinity for decays into down-type fermions. However, the
magnitude of this effect~($\simord2\%$) is relatively small,
comparable in size to the uncertainty, and might not be preserved at
higher order. Similarly, the small increase of these fermionic decay
widths as compared to their SM~counterparts is not really exploitable
at the level of the branching ratios, since the apparent constancy of
the decay widths into EW~gauge bosons is likely an artifact of the
tree-level description in these channels. The slight reduction of the
digluon decay width agrees with the impression of a somewhat reduced
coupling to top quarks, even though the large theoretical uncertainty
there forbids any definite conclusion. Similarly, little can be
deduced from the diphoton decay channel at~1L, as the large range of
variation between the~OS and original schemes~($\simord30\%$)
illustrates the significant uncertainties at this order of the
calculation. The rather clean situation in this scenario without a
light \CP-even singlet-dominated state thus demonstrates the
deterioration of the precision due to the instability of the Higgs
potential with respect to radiative corrections in SUSY-inspired
models. Working in an effective framework where the various Higgs
masses can be renormalized~OS probably improves the prediction, since
it allows to expand the perturbative series from a position in
parameter space that is closer to the physical situation, hence
promises a faster convergence of the series. On the other hand, such a
procedure explicitly violates the form of the softly-broken SUSY
potential, implying a lesser regularity of the radiative corrections.

For the scenario in the PQ-limit, the situation is complicated by the
presence of a comparatively light \CP-even singlet~$h^S = h_1$ with
tree-level mass~$\simord50$\,GeV and~$\simord90$--$100$\,GeV
at~1L. This state has a coupling to EW~gauge bosons at the level
of~$10^{-2}$ (dominated by 1L~effects) of that of a SM~Higgs at the
same mass, so that its production at~LEP is heavily suppressed. In
this scenario, the impossibility to reliably quantify the mixing with
the SM-like state~$h^{\text{SM}} = h_2$ at 1L~order is illustrated by
the wide spread of predictions in the fermionic channels,
reaching~$\simord3\%$ (for channels supposedly known with better
than~$1\%$ accuracy in the~SM, discarding the parametric
uncertainty). Here, we considered on one side the strict expansion
formalism, with the original Lagrangian, with an OS~SM-like Higgs or
with both an OS~SM-like Higgs and an OS~\CP-even singlet, on the other
side, the mixing formalism with the original Lagrangian or with an
OS~SM-like Higgs. As explained above, the OS~condition on the SM-like
Higgs mass is obtained through the introduction of a
non-vanishing~$\ell_2$. To renormalize also the \CP-even singlet
mass~OS in this scenario, we introduce an additional quartic singlet
term~$\hat{\kappa}^2\,\lvert S\rvert^4$ to the Lagrangian, together
with its
counter\-term~\mbox{$\delta\hat{\kappa}^2\approx-\hat{\kappa}^2$}, in
a very similar fashion ($A_{\kappa}$~is already needed to control the
light \CP-odd mass). This choice is natural as the mass of the
\CP-even singlet indeed proceeds mostly from such a quartic coupling
in the PQ-limit. All these operations \AP~aim at controlling the
mass-splitting and Higgs mixing already at tree level. Simultaneously,
by altering the strength of the Higgs coupling, the OS~procedures also
slightly modify the magnitude of the 1L~corrections: these effects are
\AP~legitimate, provided the added operators adequately model the
impact of the radiative corrections. In contrast, the mixing formalism
accounts for the modified mass-splitting without altering the strength
of the Higgs couplings. These differences in treatment account for the
disparity of the predictions, which can only be resolved after
inclusion of full higher-order corrections to the Higgs decays. Once
again, the 1L~EW~description of the digluon and diphoton decay widths
proves insufficient to accurately characterize these channels.

In the scenario\,CPE1 with a light \CP-even Higgs and small~$\lambda$,
the various considered approaches give very similar results in the
fermionic channels. The question of working with an SM-like
Higgs~$h^{\text{SM}} = h_2$ that is renormalized~OS or that possesses
the far-away tree-level mass of the SUSY~potential affects these decay
widths by at most a~few permil. Indeed, although the singlet--doublet
mixing is not well known in absolute value for this benchmark, its
general order of magnitude at~tree and
1L~level,~$U_{21}=\mathcal{O}(10^{-2})=\mathcal{Z}_{21}$, cannot
significantly threaten the dominant SM-like amplitude---which it
affects
as~$\simord\big(1-1/2\,\mathcal{Z}_{21}^2\big)\,\mathcal{A}^{\text{SM}}$.
The resulting effects thus remain negligible. This situation contrasts
violently with that of the singlet-dominated state~$h^S = h_1$, which
we presented in the previous section. The digluon and diphoton decay
channels still illustrate the impact of the tree-level description
inherited from the SUSY construction (which is quite different from
the physical situation in this scenario) on the precision of the
radiative widths.

In the scenario\,CPE2 with large~$\lambda$, the impact of the mixing
effect is larger than in~CPE1, and leads to a dispersion of the
predicted decay widths at the percent~level. Nevertheless, the general
order of magnitude of the decay widths for the SM-like
state~$h^{\text{SM}} = h_2$ is once again under control
(contrarily to the case of the light singlet~$h^S = h_1$).

\subsection{Interplay with the light singlets}

The presence of light-singlet states~$a^S$
  (\CP-odd) or~$h^S$ (\CP-even) modifies the phenomenology of the
SM-like Higgs~$h^{\text{SM}}$. The most obvious effect
consists in the opening of the kinematical window for the
unconventional decay into a pair of light singlets. The latter could
be sizable, in general, as the mass scales in the Higgs potential are
frequently larger than those of SM~particles. In such a case, the
properties of the SM-like state would be shadowed by the
unconventional channels, leading to inconsistency or tension with the
experimental measurements. At the technical level, the calculation of
the Higgs-to-Higgs decays raises a difficulty that is not unlike the
one discussed in \sect{sec:singletdec}. The evaluation of the
loop-functions with \texttt{LoopTools} is indeed problematic if the
tree-level Higgs masses
verify~$m_{h^{\text{SM}}}\!<2\,m_{a^S\!,h^S}$: this results
in an effective (unphysical) upper limit on the mass of the light
singlet allowing for the calculation of the decay width. An obvious
fix, similar to our strategy enabling the mass determination in
pseudo-tachyonic regions, consists in restricting ourselves to
renormalization schemes for~$A_{\kappa}$ where the considered decay is
kinematically allowed already at tree level. Below, we consider two
schemes where the singlet mass is set to~$\simord43.5$\,GeV
(scheme\,A$^{\prime}$) or~$\simord3$\,GeV (scheme\,B$^{\prime}$) at
tree level. In addition, if we allow for quartic Higgs parameters
beyond the SUSY~potential, it becomes possible to renormalize both the
light singlet and the SM-like Higgs~OS, as we discussed before, in
which case the tree-level masses trivially satisfy the kinematical
conditions already at tree level. On the other hand, the recourse to
the mixing formalism in the presence of light \CP-even
singlet-dominated states is made about impossible by the technical
difficulty in evaluating the three-point functions, as
also~$m_{h^S}\!>2\,m_{a^S\!,h^S}$ would then be needed for competitive
use of \texttt{LoopTools}.

\begin{figure}[p!]
  \centering\captionsetup{singlelinecheck=off}
  \includegraphics[width=\linewidth]{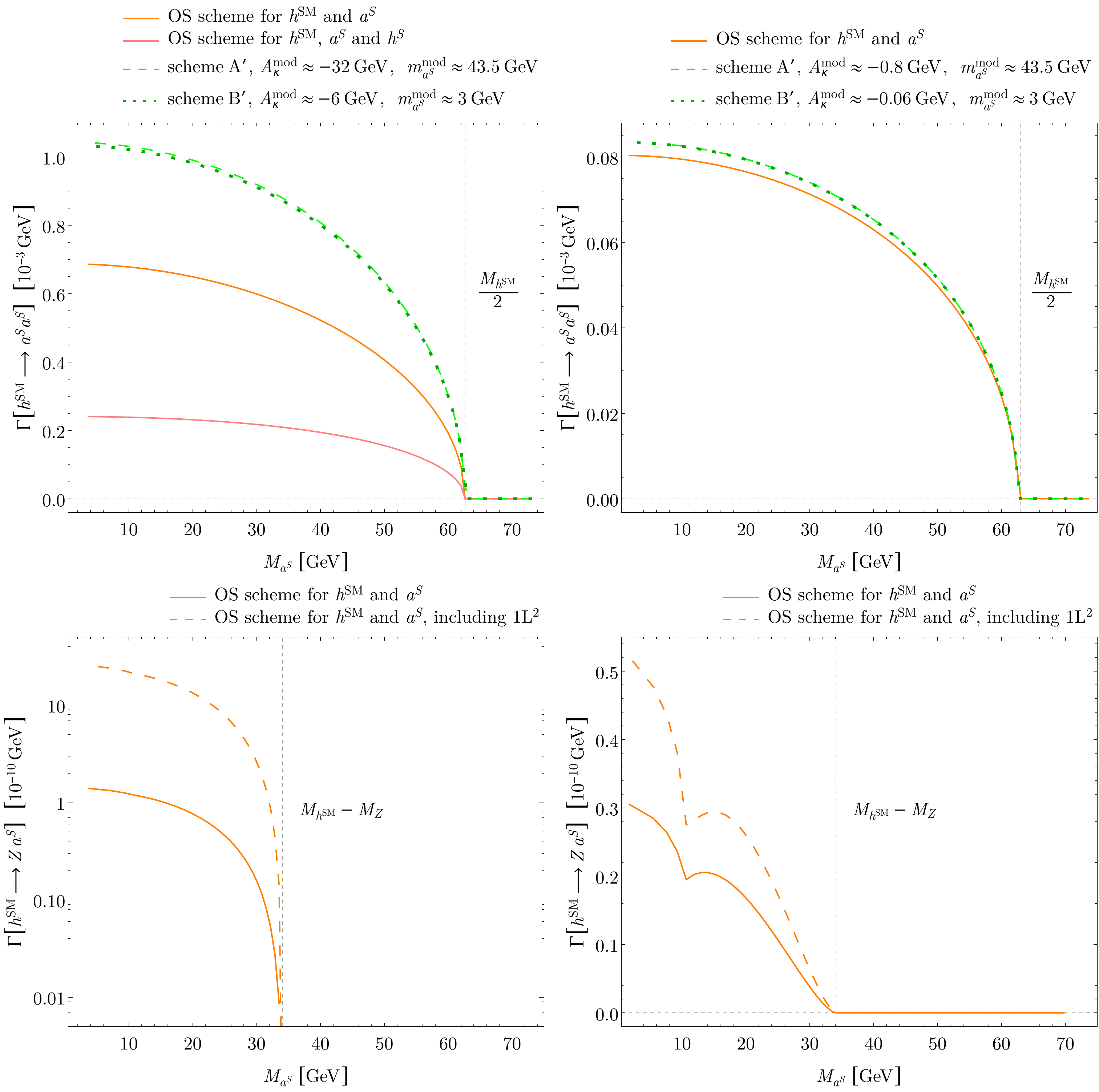}
  \caption{Predictions for the decay
    widths~$\Gamma[\AtoB{H^{\text{SM}}}{a^Sa^S}]$ (first row) and
    $\Gamma[\AtoB{H^{\text{SM}}}{Z\,a^S}]$ (second row) in the
    scenarios with light pseudo-scalars of \sect{sec:CPoddmass}:
    PQ-limit (left) and $R$-symmetric limit (right). Several schemes
    are considered:
    \begin{itemize}[leftmargin=\widthof{$\bullet$}+\labelsep]
    \item without modifying the Higgs potential, $A_{\kappa}$~is fixed
      so that the tree-level mass of the light \CP-odd state is kept
      equal to~$43.5$\,GeV (scheme\,A$^{\prime}$; dashed green curves)
      or to~$3$\,GeV (scheme\,B$^{\prime}$; dotted green curves);
    \item after introduction of a coupling~$\ell_2$ to the Higgs
      potential, the mass of the SM-like Higgs is renormalized~OS,
      while~$A_{\kappa}$ is fixed by the OS~condition on the mass of
      the light \CP-odd singlet (orange curves);
    \item in the PQ-limit, in the presence of a light \CP-even Higgs,
      the mass of the latter can also be brought~OS after introduction
      of a quartic singlet coupling~$\hat{\kappa}^2$ (pink line).
    \end{itemize}\label{fig:HSMA1A1dec}}
\end{figure}

In the scenarios with light \CP-odd Higgs~$a^S = h_4$
that we introduced in \sect{sec:CPoddmass}, the approximate symmetries
imposed on the Higgs potential constrain the triple-Higgs
couplings~$g^{\text{PQ}}_{h^{\text{SM}}a^Sa^S}$ (PQ-limit)
  and~$g^{R}_{h^{\text{SM}}a^Sa^S}$ ($R$-symmetry limit). Under the
assumption of a subleading doublet component in the light singlet and
a subleading singlet component in the SM-like state, it is possible to
derive
\begin{subequations}
\begin{align}
  g^{\text{PQ}}_{h^{\text{SM}}a^Sa^S}\! &\approx
  9\,\sqrt{2}\,\kappa\,\lambda\,v\,s_{\beta}\,c_{\beta}\,,\\
  g^{R}_{h^{\text{SM}}a^Sa^S}\! &\approx
  \frac{3\,\lambda^2\,v}{2\,\sqrt{2}\,\kappa\,\mu_{\text{eff}}}\left\{
  6\,\kappa\,s_{2\beta}\left(\frac{M_A^2\,s_{2\beta}}{2\,\mu_{\text{eff}}}
  - \frac{\kappa}{\lambda}\,\mu_{\text{eff}}\right)
  + \left(\lambda - \kappa\,s_{2\beta}\right) A_{\kappa}\right\}.
\end{align} 
\end{subequations}
It thus appears that these couplings are proportional to those
parameters that vanish in the exact $R$-symmetry or
PQ-limits. However, with our particular choice of $U(1)$-breaking
parameters, it is clear that~$g^{R}_{h^{\text{SM}}a^Sa^S}$ receives an
extra~$\lambda\,v/\mu_{\text{eff}}$ suppression
while~$g^{\text{PQ}}_{h^{\text{SM}}a^Sa^S}\!=\mathcal{O}(m_b)$:
consequently, the unconventional decay channel is potentially
important in the PQ-inspired scenario despite the coupling being
apparently protected by a symmetry. In this latter case, however, the
radiative corrections are also likely to unsettle the tree-level
picture, because of the proximity in mass of the \CP-even singlet and
the SM-like state: the assumption of a negligible singlet component in
the SM-like Higgs then appears as a fine-tuning matter and can be
challenged by mixing effects at the loop order. Unfortunately, these
are ill-controlled at~1L, as already explained before, and would
require a higher-order calculation for a quantitative estimate of
their impact on the phenomenology of the considered scenario.

In the first row of \fig{fig:HSMA1A1dec}, we consider the
Higgs-to-Higgs decay~\AtoB{h^{\text{SM}}\!}{a^Sa^S} in the two
scenarios with light \CP-odd Higgs~$a^S = h_4$, with approximate~PQ-
(left) or $R$-symmetry (right). Calculations in scheme\,A$^{\prime}$
and~B$^{\prime}$ give very similar predictions with deviations at
typically~$1\%$ (dashed and dotted green lines). 1L~effects amount
to~$\mathcal{O}(10\%)$ in the PQ-like benchmark and a~few percent in
the $R$-symmetric scenario. Therefore, the inclusion of radiative
corrections does not destabilize the tree-level amplitudes in these
schemes. Differences are more marked if we renormalize both the
\CP-odd singlet and the SM-like Higgs~OS (orange), especially in the
PQ-scenario, where the decay width is reduced. This effect is not
completely unexpected as the resummation of mass effects in the Higgs
potential simultaneously affects the Higgs-to-Higgs couplings, and the
PQ-scenario is more sensitive to radiative corrections due to the
presence of the light \CP-even singlet
at~$\simord100$\,GeV. Renormalizing also the mass of the latter~OS
(pink curve) further widens the scope of predictions for the
Higgs-to-Higgs decay. We must thus conclude that the inclusion of
higher orders is needed to control the uncertainties of about~$100\%$
for this channel in the PQ-inspired scenario---while the situation is
comparatively tame in the scenario with approximate $R$-symmetry. Now
looking at the magnitude of the
predicted~\AtoB{h^{\text{SM}}\!}{a^Sa^S} decay width, we observe that
it remains below~$2\%$ of the total width of the SM~Higgs, in the
$R$-symmetric scenario, thus defying the sensitivity of detection at
the~LHC. In the PQ-scenario,
however,~$\Gamma[\AtoB{h^{\text{SM}}\!}{a^Sa^S}]$ could
reach~$\simord1$\,MeV at low~$M_{a^S}$, leading to tensions with the
measured Higgs properties. Nevertheless, as we stressed above, the
large theoretical uncertainty does not allow to unquestionably
associate this phenomenology to the \mbox{investigated point in
  parameter space}.

Another possible decay channel is~\AtoB{h^{\text{SM}}\!}{Z\,a^S}. In
this case, given that~$m_{h^{\text{SM}}}\!\lsim M_Z$ at tree level,
the difficulty of \texttt{LoopTools} in evaluating loop functions
becomes critical, and the only consistent mean at our disposal for
computing the decay width (without actually repairing the evaluation
of three-point functions) consists in renormalizing the mass
of~$h^{\text{SM}}$~OS through the introduction of a
non-vanishing~$\ell_2$ in the Higgs potential. The results are shown
in the second row of \fig{fig:HSMA1A1dec} (solid orange curves) and
the predicted widths are clearly subleading. In the scenario with
approximate PQ-symmetry, we may also renormalize the mass of the light
\CP-even singlet~OS via the introduction of the quartic
parameter~$\hat{\kappa}^2$: corresponding results are
indistinguishable from those where the mass of~$h^S$ is computed in
the original scheme (so that we do not present the corresponding
results in the plot). On the other hand, the 1L amplitude dominates
the process, as is demonstrated by the inclusion of the 1L$^2$~term
(dashed orange curves).  This shows that the decay width is controlled
by effects of 2L~order. The situation is less critical in the scenario
with approximate $R$-symmetry (right-hand side) but the
1L$^2$~contribution is still comparable to the strict 1L~width,
hinting at the need to include 2L~corrections for a reliable control
of the decay width.

At the 1L~order, also the tree-level three-body
decays~\AtoB{h^{\text{SM}}\!}{a^Sf\bar{f}}, with~$f\in\{b,\tau\}$, are
\AP relevant\,\cite{Domingo:2019vit}. However, our results showed no
significative contribution at the numerical level beyond that of
the~OS~\AtoB{h^{\text{SM}}\!}{a^Sa^S}~channel. This is consistent with
the absence of a hierarchy between the masses of the decaying Higgs
and of the EW~gauge bosons, forbidding the emergence of large
Sudakov~logarithms. The PQ-inspired scenario also contains a
relatively light singlino-dominated state~$\chi_1^0$ with
mass~$\simord43$\,GeV, allowing for the \AP~invisible decay
channel~\AtoB{h^{\text{SM}}\!}{\chi_1^0\,\chi_1^0}: its width is
at~$\mathcal{O}{(1\,\text{MeV})}$, and depends on the details of the
singlet--doublet mixing. We do not investigate it further in this
article, as such channels are not the focus of our discussion.

\begin{figure}[b!]
\centering
\includegraphics[width=\linewidth]{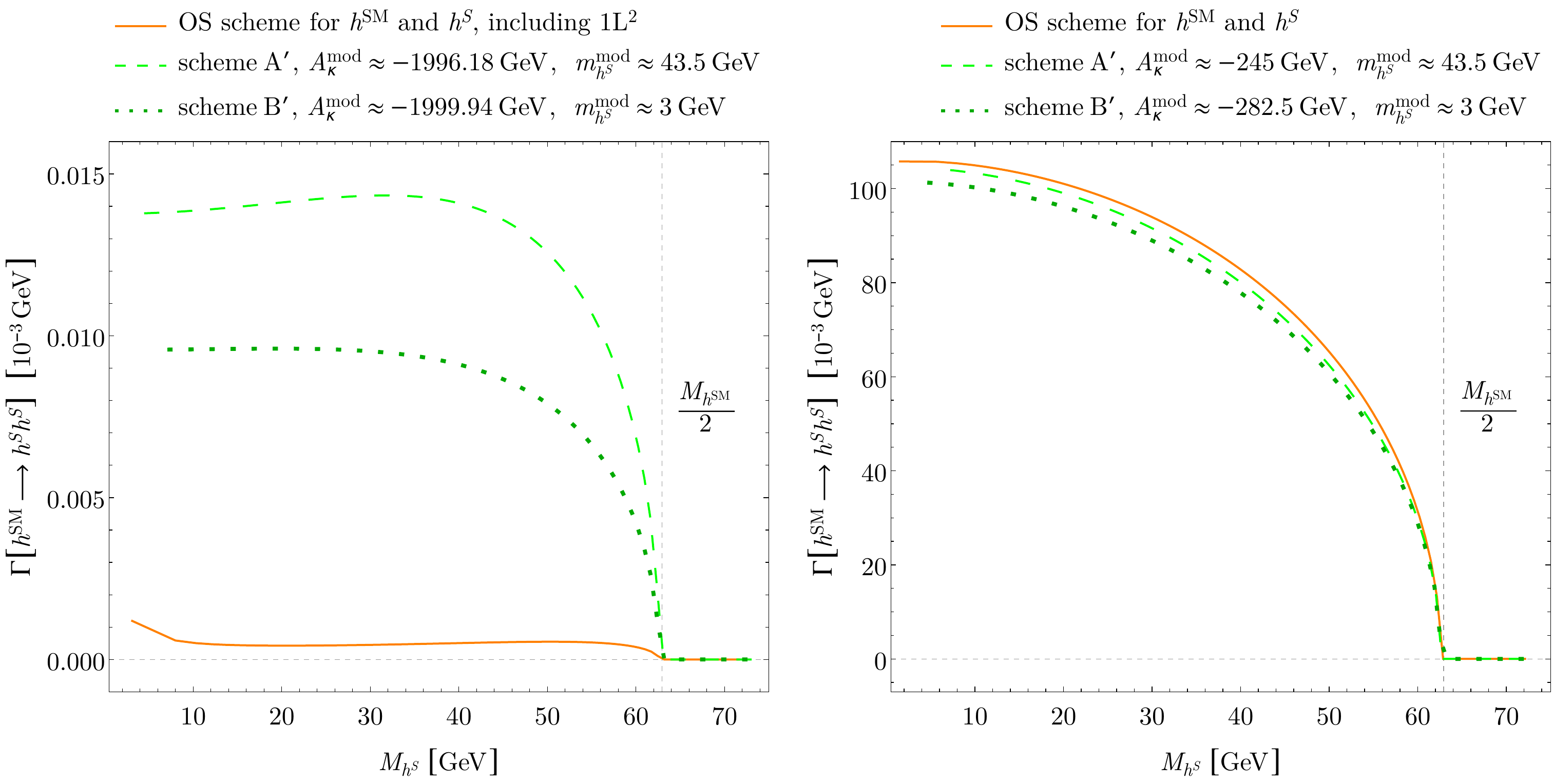}
\caption{Predictions for the decay
  widths~$\Gamma[\AtoB{h^{\text{SM}}\!}{h^Sh^S}]$ in the scenarios
  with light scalars of \sect{sec:CPevenmass}: CPE1 with low value
  of~$\lambda$ (left), CPE2 with large value of~$\lambda$ (right). The
  considered schemes are comparable to those of
  \fig{fig:HSMA1A1dec}. \label{fig:CPEHhh}}
\end{figure}

\needspace{9ex} In the scenarios of \sect{sec:CPevenmass}, with a
light \CP-even state~$h^S = h_1$ and no symmetry protecting the Higgs
sector (or a symmetry that is broken at a scale comparable to the
EW~one), the reliability of the predictions for the Higgs-to-Higgs
decays at~1L can be expected to be poor, similarly to what was
observed for the singlet--doublet mixing in \sect{sec:CPEdec}. The
corresponding results are displayed in \fig{fig:CPEHhh}. The plot on
the left shows that, in the scenario\,CPE1 with small~$\lambda$, the
choice of scheme leads to predictions
for~$\Gamma[\AtoB{h^{\text{SM}}\!}{h^Sh^S}]$ ranging from negligible
values to percent~level of the SM~total width below the kinematical
threshold. This suggests in any case that the unconventional Higgs
decay would be difficult to identify at the~LHC. In the plot on the
right, the scenario\,CPE2 with large~$\lambda$ leads, under threshold,
to large decay widths, incompatible with the measured Higgs
properties. In this case, the Born-level amplitude dominates the
process and sets the general scale of the decay width, although the
variation induced by the choice of scheme suggests an uncertainty
of~$\mathcal{O}(10\%)$. \CP-conservation forbids
the~\AtoB{h^{\text{SM}}\!}{Z\,h^S}~decay in these scenarios.

As a last comment about the limited precision achieved at~1L in the
prediction of these Higgs-to-Higgs decays, we observed that variations
on the definition of the quark masses (\AP a~2L~matter) were liable to
produce shifts of~$\mathcal{O}{(100\%)}$ at the level of the widths,
\EG~in scenario\,CPE1, emphasizing the large uncertainty which we
illustrated above only through alternating descriptions of the
tree-level Higgs spectra.

\section{Conclusions \label{sec:concl}}

In this paper, we have shown how apparent tachyonic masses in the
tree-level spectrum of the NMSSM could (sometimes) be seen as an
artificial problem related to an inadequate choice of renormalization
scheme. In the specific case of the $Z_3$-conserving NMSSM with a
light \CP-even or \CP-odd singlet-dominated state, the renormalization
of $A_{\kappa}$ is a natural handle on such a mass, and generally
allows for the selection of a scheme with a well-defined tree-level
spectrum. An OS renormalization of the light state is even possible,
and may prove a viable strategy in order to resum logarithms caused by
such a light-singlet mass.

The choice of a renormalization scheme with well-defined tree-level
spectrum also allows to calculate further physical properties
of the light Higgs bosons, in particular their decay widths or their
impact on the phenomenology of the SM-like state. There, we showed
several additional difficulties that the mismatch between tree-level
and physical spectra could cause, in particular the occasional
emergence of three-point functions corresponding to `unphysical'
configurations (\IE~with the sum of masses of the decay
products exceeding the one of the decaying
particle at tree level) or the slow convergence of the perturbative
series in the presence of a light singlet-dominated \CP-even state
(causing masses, mixing matrices and couplings to take tree-level
values very far from the actual situation). The use of
OS~renormalization conditions and the introduction, if needed, of
additional non-SUSY parameters to the Higgs potential often alleviate
these problems, or at least shed some light on the uncertainties at
stake.

The scenarios that we considered in this paper are strictly meant as
illustration of the computational issues that we wished to discuss and
should not be understood as presenting a comprehensive picture of the
NMSSM phenomenology in light-Higgs scenarios. Nevertheless, it is
tempting to emphasize a few general features of such physical
configurations. The main two effects are the opening of new decay
channels for the SM-like state and, if the light state is \CP-even,
the exchange of singlet and doublet properties via mixing. We have
stressed, however, how these two features were not necessarily so well
controlled at the considered orders achieved in the calculation.

\needspace{3ex}
The theoretical sources of uncertainty are necessarily more numerous
in BSM models than in the SM. In this respect, we have discussed how
the comparatively large difference between tree-level and physical
mass for the SM-like state of the SUSY model could amount to a few
permil uncertainty in fermionic decays at~NLO and up to~$\simord30\%$
for the digluon and diphoton decays at (QCD-corrected)~LO. While we
believe that this feature is improved by the inclusion of additional
quartic couplings in the Higgs potential, allowing for an
OS~renormalization of the relevant masses (but departing from the SUSY
framework), this gain for the convergence of the perturbative series
is only likely to be verifiable after the actual inclusion of
higher-order corrections to the decay widths. In addition, reliable
predictions at the level of the branching ratios require homogeneous
control over all the leading decay channels, failing which, the
uncertainty of the less accurate widths contaminate all other
channels. In the case of SM-like states in SUSY models, this suggests
going beyond the approximation of
off-shell~\AtoB{h^{\text{SM}}\!}{WW^*,ZZ^*} at tree-level. Finally,
the lack of control over the singlet--doublet mixing in scenarios with
a light \CP-even singlet could seriously hinder the predictivity of
calculations at~1L, both at the level of the properties of the light
Higgs and in assessing the deviations with respect to the~SM for
widths of the SM-like state (in scenarios with large~$\lambda$). There
again, only higher orders could satisfactorily improve the reliability
of the predictions and prepare such scenarios with light
singlet-dominated states to a meaningful comparison with precise
experimental measurements, as expected from future~$e^+e^-$~colliders.

\section*{\tocref{Acknowledgments}}

F.\,D.~acknowledges support of the BMBF Verbund-Projekt 05H2018 and
the DFG grant SFB CRC-110. The work of S.\,P. is supported by the BMBF
Grant No. 05H18PACC2.

\appendix

\needspace{51ex}
\section{Input\label{app:in}}
\begin{table}[h!]
\centering\renewcommand{\arraystretch}{1.5}
\begin{tabular}{|r@{\,}l | c | c | c | c |}
\hline
& & $R$-symmetric & PQ-limit & CPE1 & CPE2 \\\hline
$\lambda$ & $\big(\DR\big)$ & $0.2$ & $0.4$ & $0.05$ & $0.5$ \\\hline
$\kappa$ & $\big(\DR\big)$ & $0.4$ & $-0.02$ & $0.05$ & $0.05$ \\\hline
$M_{H^{\pm}}$ & $\big(\text{OS}\big)$\, [TeV] & $1.75$ & $2.5$ & $4.8$ & $1.225$ \\\hline
$t_{\beta}$ & $\big(\DR\big)$ & $10$ & $5$ & $10$ & $2$ \\\hline
$\mu_{\text{eff}}$ & $\big(\DR\big)$ [GeV] & $400$ & $-478.5$ & $500$ & $500$ \\\hline
$A_{\kappa}$ & $\big(\DR\big)$ [GeV] & $[-2,0.5]$ & $[-60,60]$ & $[-2003,-1990]$ & $[-500,-380]$ \\\hline
$A_{t}$ & $\big(\DR\big)$ [TeV] & $-2.5$ & $-2.5$ & $-2.4$ & $2.1$ \\\hline
\end{tabular}
\caption{NMSSM input for the various scenarios considered in this
  paper. The remaining parameters, $M_1=450$\,GeV, $M_2=700$\,GeV,
  $M_3=2$\,TeV, and the quadratic soft SUSY-breaking terms governing
  the sfermion masses,~$m_{\tilde{f}}=1.5$\,TeV, are not varied. The
  renormalization scale for \DR~parameters
  is~$m_t$.\label{tab:NMSSMinput}}
\end{table}

\needspace{20ex}
\bibliographystyle{h-physrev}
\bibliography{literature2}

\end{document}